\documentclass[preprint,aps,showpacs,nofootinbib]{revtex4-1}
\pdfoutput=1 
\usepackage{amsmath}
\usepackage{amsfonts}
\usepackage{amssymb,amscd,amsbsy}
\usepackage{color}
\usepackage{graphicx}
\usepackage{subcaption}

\usepackage{mathtools}

\newcommand{\bea}{\begin{eqnarray}}
\newcommand{\eea}{\end{eqnarray}}

\newcommand{\vect}[1]{\mathbf{#1}}

\newcommand{\tr}{\textcolor{red}}

\newcommand{\kt}{k_{\rm B}T}

\newcommand{\cref}{c^{(2)}_{\rm ref}}

\begin{document}
\title{Nematic and gas-liquid transitions for sticky rods on square and cubic lattices}

\author{P. Quiring, M. Klopotek and M. Oettel}
\email{martin.oettel@uni-tuebingen.de}
\affiliation{
Institut f\"ur Angewandte Physik, Eberhard Karls Universit\"at T\"ubingen, D--72076 T\"ubingen, Germany \\
}

\begin{abstract}
Using grand--canonical Monte Carlo simulations, we investigate the phase diagram of hard rods of length $L$ 
with additional contact (sticky) attractions on square and cubic lattices. 
The phase diagram shows a competition between gas--liquid and ordering transitions (which are
of demixing type on the square lattice for $L \ge 7$ and of nematic type on the cubic lattice for $L \ge 5$). 
On the square lattice,
increasing attractions initially lead to a stabilization of the isotropic phase. On the cubic lattice,
the nematic transition remains of weak first order upon increasing the attractions. In the vicinity
of the gas--liquid transition, the coexistence gap of the nematic transition quickly widens. These features
are different from nematic transitions in the continuum.  
\end{abstract}

\maketitle

\section{Introduction}
\label{sec:intro}

The study of phase transitions in fluids with anisotropic particles has been of great interest in
the statistical physics and soft matter communities for the past decades. From the theory side, it all
started with Onsager's seminal study on the existence of a first--order transition between an isotropic and a nematic state in a hard--rod fluid in the limit of infinitely long rods \cite{Onsager1949}. Meanwhile, the phase behavior of hard rods in the continuum is fully established \cite{Bolhuis1997}. Nematic ordering starts from a critical aspect ratio $a$ (ratio between length and diameter) of about 3.8. Symmetry arguments and mean--field
theory predict the transition to be first order, but an associated density gap is hard to detect near the critical aspect ratio. It is visible for $a=5$ \cite{Dijkstra2006} and reaches
the Onsager limit already for $a \gtrsim 15$  \cite{Bolhuis1997}.

The case of rods with additional attractions has been studied mainly in the context of the Asakura--Oosawa (AO) model, appropriate for systems of hard rods where the addition of non--adsorbing polymers provides tunable attractions through the depletion effect. Here, the concentration of polymers corresponds to an inverse temperature. Mean--field theories such as free volume theory for the AO model of rods and polymers \cite{Lekkerkerker1994,Lekkerkerker1997} predict a continuous and substantial widening of the isotropic--nematic density gap. In the case of small polymers (corresponding to short--ranged attractions), the transition smoothly crosses over to a transition between an isotropic gas and a nematic liquid. This scenario is basically confirmed in simulations, but, the nematic transition becomes metastable with respect to the isotropic gas--crystal
transition in those cases \cite{Lekkerkerker1997,Dijkstra2006}. The widening of the coexistence density gap has been observed  experimentally in mixtures of fd--virus and dextrane polymer mixtures \cite{Fraden2004} and in ones of boehmite rods and PDMS or PS polymers \cite{Lekkerkerker1995}, although arrested states appearing further complicates matters.

In the theoretical exploration of phase transitions, in general, the study of simple lattice models has often helped to investigate generic features. This has been the case for the ordinary gas--liquid transition whose lattice counterpart is the lattice gas (or Ising) model. 
As argued, phase transitions in systems where particles have anisotropic  interactions is of fundamental interest for systems with continuous orientation--degrees--of--freedom. 
For 2D systems,  the Lebwohl--Lasher / XY--models \cite{Leb1972,Kost1974,Mol2006} have early taken on the role as basic models for phase transitions in liquid crystals. Also, the Potts model in 2D (with discrete spin--orientation degrees of freedom) \cite{Wu1982} may serve as an approximate analogy to liquid crystals in 2D \cite{Vink07}. These all share one common feature: They are \emph{on--lattice} and have anisotropic potential--type interactions that do \emph{not} stem from local, steric-repulsion due to particle shape. On the other hand, the study of `soft matter' systems, which are almost always treated in the off--lattice context, have long incorporated anisotropic--particle models.  
Therefore, we think that on--lattice models of \emph{sterically} anisotropic particles  complement both existing soft--matter-- and idealized lattice--spin--type treatments of statistical systems.

On--lattice hard rods could possibly serve as an alternative workhorse model.
One may ask whether they are likewise generic for nematic systems. For rods on a cubic lattice, say, positional as well as orientational degrees of freedoms are discretized. An essential difference, however, is that by simply making the lattice ``denser'' there is no  continuum limit possible for the orientational degrees of freedom. The continuum limit is actually the so--called Zwanzig model \cite{Zwanzig1963} with continuous positional degrees of freedom and 3 possible orientations of the rods in the Cartesian directions.  Such long, hard rods render virial coefficients that scale very differently with respect to their aspect ratio from those of hard--rod models with unrestricted orientations. Therefore, a second--virial approximation is not sufficient to locate the isotropic--nematic transition even in the limit of long rods. A density-functional treatment of the Zwanzig model for hard rods in three dimensions (3D) using fundamental measure theory (FMT) shows many qualitative similarities in the phase diagram generated compared to  the one for continuum rods \cite{MartinezRaton2004}, but, additional phases appear. Simulation results for the phase diagram of the continuum Zwanzig model in 3D are not known to us, neither are studies on the influence of additional attractions between the rods. 

The nature of continuous symmetries in  two-dimensional (2D) systems can render ordering transitions that are quite unique, specifically for those of the Kosterlitz--Thouless--type. One may then ask what the 2D situation is for rod and liquid crystal systems. As a first point, simulations for the 2D Zwanzig model have shown that the ordering transition is one where $x$-- and $y$--oriented rods \emph{demix} with the same character of the gas--liquid transition in the lattice gas (Ising universality class) \cite{Vink2009}. This is not a nematic transition, in contrast to the 3D case. Continuous, 2D liquid--crystal models display an isotropic--\emph{quasinematic} transition ( for needles \cite{Fre85,Vink14} and ellipses \cite{Fre90,Xu13}), where, however, the order is not truly long--ranged. This can even turn into a first--order transition \cite{Fre90,Vink07,Vink14}. Ref. \cite{Fre85} discusses how an isotropic--quasinematic transition can indeed be of Kosterlitz--Thouless type, shown for hard ellipses in Ref. \cite{Fre90} and needles in Refs. \cite{Barma05,Vink14}.

This discussion on possible types of ordering transitions points to a general difference between rod models on the lattice and in the continuum. Previous studies of hard rods with unit square cross section
on a cubic lattice in 3D \cite{Gschwind2017,Rajesh2017} have detected a nematic transition for $a \ge 5$. In stark contrast to continuum models, the nematic phase can be manifested in a  peculiar manner, where there is a `negatively--preferred' direction instead. Rods oriented in this direction
are suppressed and the system splits into weakly--correlated, effectively  two--dimensional layers within which rods are distributed isotropically. This version of a nematic phase is the stable one for $a=5$ and 6. Furthermore, the nematic transition is of very weak first order for the aspect ratios studied (up to 25) \cite{Gschwind2017}.  A coexistence density gap is not detectable and the first order character is only visible through a finite--size--scaling analysis for large lattices. Analytic results, in constrast, predict a strong first order transition for $a \ge 4$ \cite{DiMarzio1961,Alben1971,Stilck2011,Oettel2016}.
The effect of additional attractions has been explored by density functional theory (FMT) \cite{Mortazavifar2017} and yields phase diagrams qualitatively very similar to those in the continuum (i.e. it reproduces a continual, strong widening of the coexistence gap with 
increasing attractions).

In this paper, we study the effect of attractions for rods on lattices by simulations in the grand--canonical ensemble. The paper is structured as follows: In Sec.~\ref{sec:model} we introduce the model and briefly describe the methods of analysis. In Sec.~\ref{sec:2D} we revisit the case of 2D, where, for the case of purely hard--core rods, the demixing transition has been studied extensively \cite{Ghosh2007,Ramirez2008,Longone2012,Kundu2013,Kundu2014,Kundu2015}. 
Upon adding attractions, the demixing transition competes with the liquid--vapor transition, yet the
full phase diagram in the temperature--density plane had not been resolved \cite{Longone2010} up to now.            
In Sec.~\ref{sec:3D} we study the 3D case for aspect ratios 4,5,6 and 8, and in Sec.~\ref{sec:summary}
we present a summary and conclusions.

\section{Model and methods}
\label{sec:model}

We consider quadratic lattices in 2D and cubic lattices in 3D, where the unit cell length is set to 1. 
Hard rods are parallelepipeds with extensions $L \times 1 $ (in 2D) or $L \times 1 \times 1$ (in 3D)  and are defined by $L$ consecutively-covered  lattice points in one Cartesian direction. We define a binary occupancy field $O({\vect s})$, whose values are 1 for lattice points ${\vect s}$ covered by rods and 0 otherwise.
The position of a rod is specified by the lattice point it covers having minimal coordinates (in each dimension). The rods are forbidden to overlap, i.e. there is no double occupancy of a lattice point. 
The quadratic lattice restricts the number of possible orientations to two, the cubic lattice to three, 
and we refer to rods oriented in $x$-- resp. $y$-- resp. $z$--direction as species 1 resp. 2 resp. 3. 
Species densities $\rho_i$ are defined as number of rods of
species $i$ ($N_i$) per lattice site, $\rho=\sum_{i=1}^D \rho_i$ is the total density, 
$\eta=\rho L \le 1$ is the total packing fraction and $D$ is the number of dimensions. 

\begin{figure}
 \centerline{\includegraphics[scale=0.5]{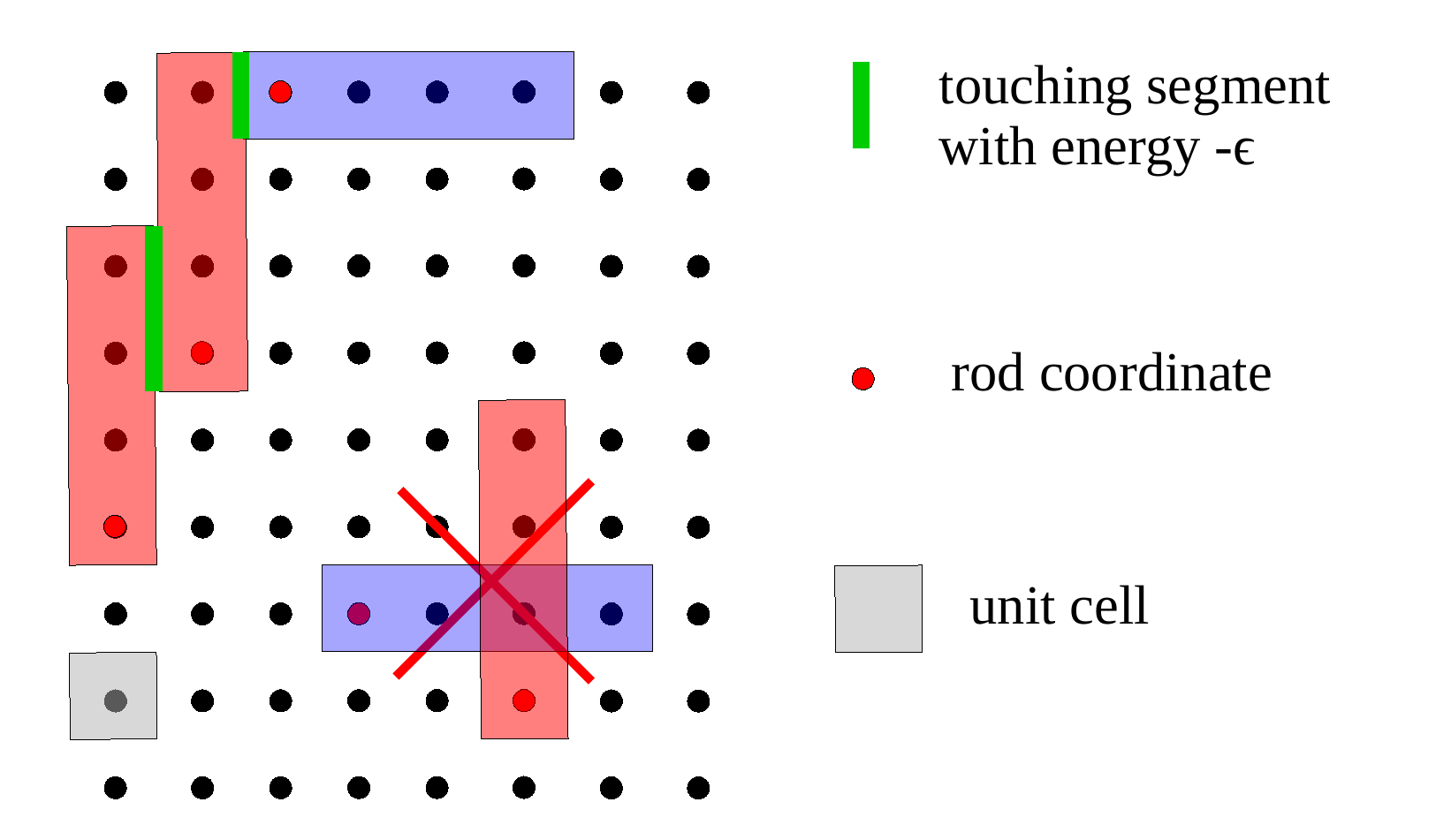}}
	\caption{2D scheme of the model. 
The common surface of touching segments of neighboring rods are shown in green.}
 \label{fig:2D_model_schematics}
\end{figure}

We consider sticky attractions between rods with an energy $-\epsilon$ $(\epsilon>0)$ per touching segment
of neighboring rods (see Fig. \ref{fig:2D_model_schematics}). 
The internal energy of a non--overlapping (valid) rod configuration $\omega$ can therefore be written as 
\bea
   U(\omega) =  -\epsilon \sum_{\langle \vect s \vect s' \rangle} O(\vect s) O(\vect s') + N(L-1) \epsilon\;,
\eea
where the sum is over neighboring site--pairs $\left\langle \vect s, \vect s'\right\rangle$ of the lattice and contributes whenever both sites are occupied. The second term corrects the over--counted adjacent sites \emph{within} each rod ($N$ is the total number of rods). With these ingredients, the grand partition function of the system is defined by
\bea
  \Xi (z_i) = \sum_{N_1=0}^\infty .. \sum_{N_D=0}^\infty \; \prod_{i=1}^D \frac{z_i^{N_i}}{N_i!} \; \sum_{\omega} e^{-\beta U(\omega)},
  \label{eq:grand_partition_function}
\eea
where $z_i =\exp(\beta\mu_i)$ is the activity of species $i$ (with inverse temperature $\beta=1/(\kt)$  and  chemical potential $\mu_i$ of species $i$). $\sum_{\omega}$ represents the sum over all non--overlapping configurations with $N_i$ rods. In a bulk system, all $z_i$ are equal ($z_i=z$) and the grand partition function can be written in a single--component form as
\bea
  \Xi (z) = \sum_{N=0}^\infty \; \frac{z^{N}}{N!} \; \sum_{\tilde{\omega}} e^{-\beta U(\tilde{\omega})}\;.
  \label{eq:grand_partition_function_simple}
\eea
$\sum_{\tilde{\omega}}$ is now the sum over all non--overlapping configurations with $N=\sum _{i=1}^D N_i$ rods. This is the considered case in this work. The phase diagram of the model is spanned by the total packing fraction $\eta$ and the reduced
temperature $T^*=\kt / \epsilon$.

\subsection{Order parameters}

In the 2D case, we introduce the normalized order parameter
\bea
  S &=  &  \frac{\rho_1 -\rho_2}{\rho}
\eea
measuring the demixing of rods oriented in $x$-- and $y$-- direction. Note that demixed states with $S$ and $-S$ are
equivalent since exchanging rods with $x$ and $y$ orientation is a symmetry of the model.
The strength of demixing is characterized by $\left|S\right|$ and will be used below 
to determine the transition between isotropic and demixed states.

\begin{figure}
 \centerline{\includegraphics[scale=0.6,angle=-90]{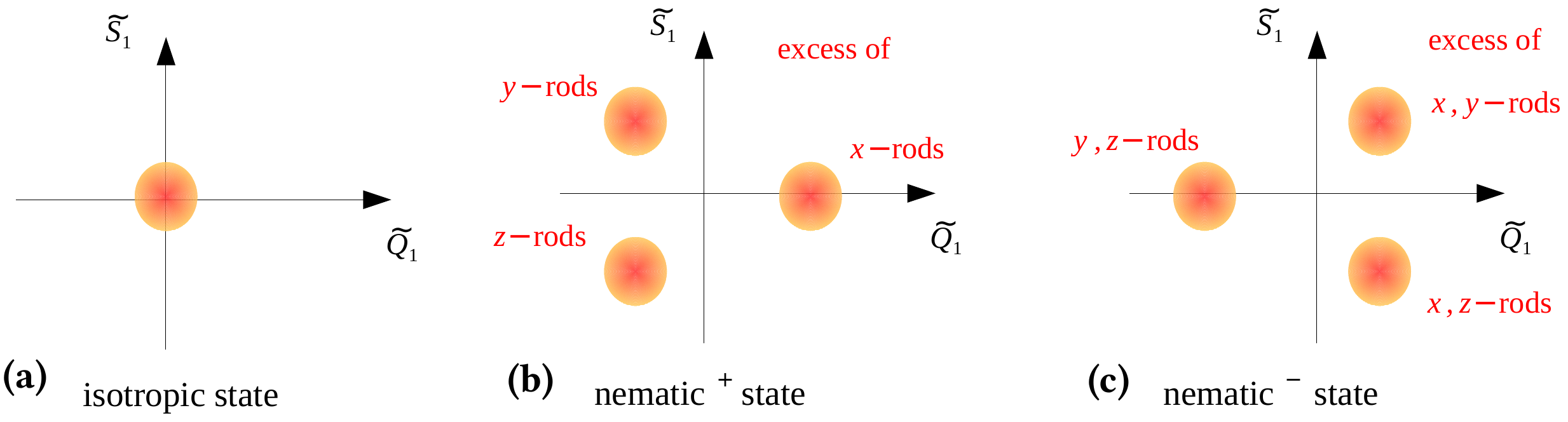}} 
 \caption{Schematic histograms for isotropic and nematic states in the $\tilde Q_1$--$\tilde S_1$ order parameter plane, 
as expected for a finite system in the grand canonical ensemble.}
 \label{fig:scheme}
\end{figure}

More symmetries exist in the 3D case: any permutation of the rod orientations renders an equivalent state. In this respect, our lattice model mimicks a three--state Potts model where the symmetry--broken phase consists of three equivalent states (each of the Cartesian axes can be the preferred (or negatively-preferred) direction in the demixed phase).
Following Ref.~\cite{Vollmayr1993}, the order parameter dimensionality must be two, and 
orthogonal axes in order parameter space are formed by pairs $(\tilde Q_i, \tilde S_i)$ of unnormalized nematic and biaxial
order parameters:
\bea
 \label{eq:tildeQidef}
   {\tilde Q_i} & = &  {\eta_i - \frac{\eta_j +\eta_k}{2}} \;, \\
 \label{eq:tildeSidef}
   {\tilde S_i} &=  &  \frac{\sqrt{3}}{2}(\eta_j -\eta_k) \;,
\eea 
where $(ijk)$ is a cyclic permutation of $(123)$. States of the system can be represented in the 
$\tilde Q$--$\tilde S$ order parameter plane. Fig.~\ref{fig:scheme} shows schematic probability 
distributions (histograms) of bulk states in the order parameter plane as expected in a finite system in
the grand canonical ensemble.  An isotropic state is given by a peak in the origin.
A nematic state with one majority species and two alike minority species  (no biaxiality) renders
three equivalent states, which arrange in a triangle pointed towards the right on the $\tilde Q_i$--axis in the order-parameter plane. This state will be
termed the ``nematic$^+$'' state. A nematic state
with one minority species and two alike majority species  renders a such triangle
pointed towards the left (``nematic$^-$'' state). 
The hypothetical case of a  nematic state with \emph{nonzero} biaxiality would render six equivalent states. We have not found 
stable biaxial states in this work, however.  

The overall degree of nematic order can be captured by the scalar order parameter
 
\bea
  m = \sqrt{\eta_1^2 + \eta_2^2 + \eta_3^2 - \eta_1\eta_2 -\eta_1\eta_3-\eta_2\eta_3}
\eea
for which $m^2 = \tilde Q_1 ^2 + \tilde S_1^2=\tilde Q_2 ^2 + \tilde S_2^2=\tilde Q_3 ^2 + \tilde S_3^2$ holds.

\subsection{Simulations}

\begin{figure}
\centering
\includegraphics[scale=0.4]{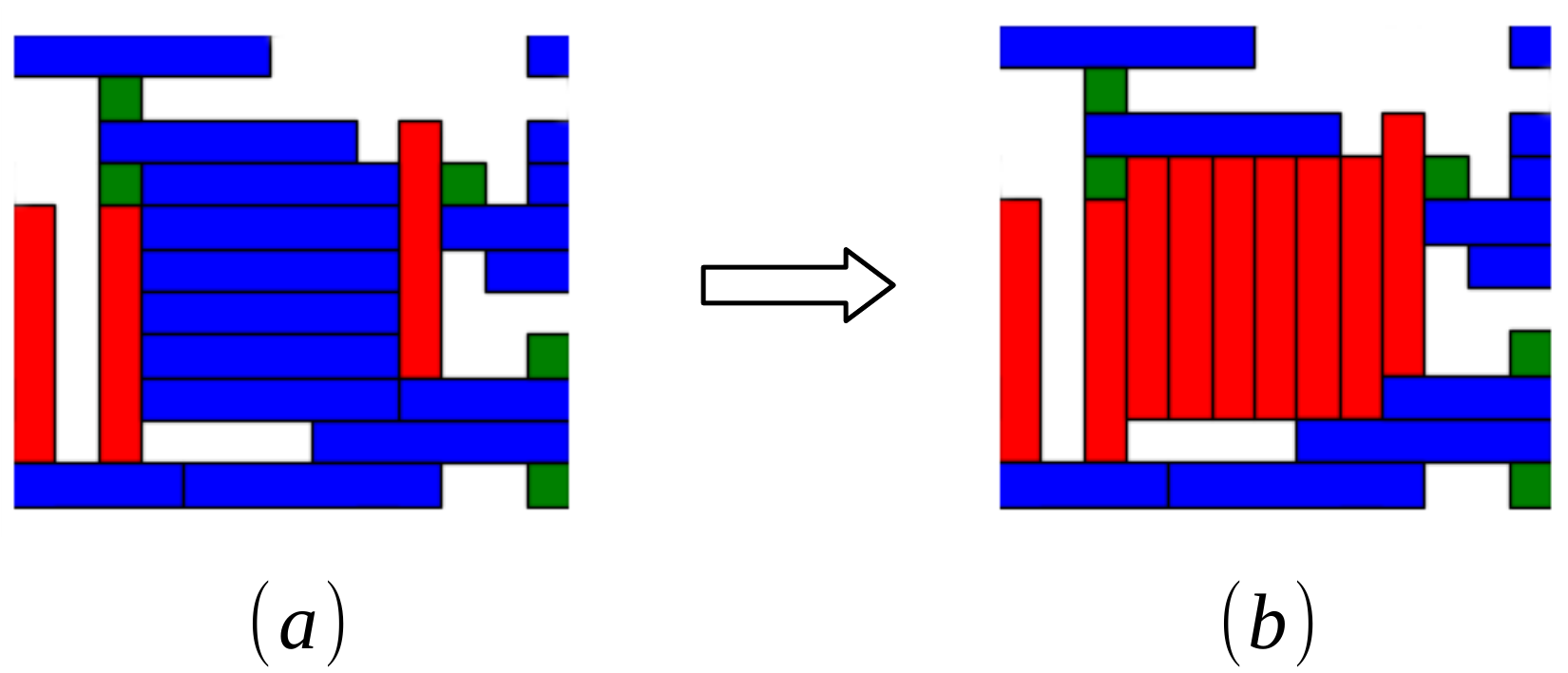}
\caption{Flip move of a $6 \times 6$ square of rods oriented in $x$--direction (a) to a $6 \times 6$ square of rods oriented in $y$--direction (b). The green squares are $z$-rods oriented perpendicular to the $xy$--plane.}
\label{fig:flip_move}
\end{figure}

The grand--canonical average of an observable $A(\tilde\omega)$ is defined by
\bea
  \langle A \rangle  = \frac{1}{\Xi} \sum_{N=0}^\infty \frac{z^N}{N!} \sum_{\tilde\omega} A(\tilde\omega)e^{-\beta U(\tilde\omega)}  \;
\eea
and is computed by a standard grand--canonical Monte--Carlo simulation algorithm
in a quadratic or cubic simulation box with linear length $M$. 
The acceptance probability of the algorithm for inserting a rod is denoted by $\alpha_{\text{ins}}$
(``$\alpha_{N \rightarrow N+1}$'') and for deleting a rod by $\alpha_{\text{del}}$ (``$\alpha_{N \rightarrow N-1}$'').
They are given by
\bea
\alpha_{\text{ins}} &= \min \left(1, \frac{\pi_{ N+1 \rightarrow N}}{\pi_{ N \rightarrow N+1}} \frac{z}{N+1} e^{-\beta(U(\tilde\omega_{N+1})- U(\tilde\omega_{N})} \right) \\
\alpha_{\text{del}} &= \min \left(1, \frac{\pi_{ N-1 \rightarrow N}}{\pi_{ N \rightarrow N-1}} \frac{N}{z} e^{-\beta(U(\tilde\omega_{N-1})- U(\tilde\omega_{N})} \right),
\eea
where $\pi$ is a proposal probability for inserting ($N \rightarrow N+1$) or deleting  a rod
($N \rightarrow N-1$). The ratio of the proposal probabilities in $\alpha_{\text{ins}}$ is equal to $DM^D$ and in $\alpha_{\text{del}}$ equal to $1/(DM^D)$. 
An additional ``flip--move'' is implemented: after a successful insertion of a rod into the lattice we check 
if the inserted rod is part of an $(L\times L)$--square fully covered by $L$ parallel rods. 
The orientations of all rods within such a square  are flipped in--plane
randomly (see fig. \ref{fig:flip_move}). Since there is no change in internal energy, detailed
balance holds. The flip move improves the performance of the algorithm at high densities.


In general, phase transitions in this system are accompanied by a change in packing fraction and in order
($|S|$ or $m$). We pinpoint the transition by the location of sharp peaks in
var($\eta$) and var($|S|$), var($m$) in the $z$--$T^*$ plane. Here, var denotes the variance $\text{var}(A)=\langle A^2 \rangle - \langle A\rangle^2$ of an observable $A$. If a transition is sufficiently strongly first order,
we pinpoint the transition by the location of the two peaks in $P(\eta)$ having equal area,
where $P(\eta)$ is the histogram (probability distribution) of packing fractions for given $z,T^*$.
This also renders the packing fractions of the coexisting states.
For low $T^*$ (strong attractions), we employ successive umbrella sampling (SUS) \cite{Virnau2004}. 

We performed a finite--size scaling (FSS) for selected systems to determine the location of
critical transitions more precisely. For this, the Binder cumulant $K_2(A)=\langle A^2 \rangle / \langle |A| \rangle^2$
of an order parameter $A$ is plotted for a few system sizes $M$ over a temperature domain near the
critical temperature $T^*_c$. The latter is then determined by the crossing point of the cumulants
\cite{BinderHeermann}.

\section{Two Dimensions}
\label{sec:2D}

The 2D system on square lattices has enjoyed some attention, especially the case of 
hard rods without attractions. The initial focus was on the onset of the
second--order transition from
an isotropic to a demixed state where one of the two rod species dominates.
This demixing is found for $L \ge 7$  in simulations~\cite{Ghosh2007}. 
The critical packing fraction for the onset of demixing scales approximately
as $4.8/L$ for large $L$~\cite{Ramirez2008,Kundu2015}.
At very high packing fractions $\eta \approx 1$, theoretical arguments predict a re--entrant
transition from the demixed to a disordered state bearing some characteristics of a 2D cubatic phase
on a lattice \cite{Ghosh2007}. 
This transition has been studied in more detail using simulations in Refs.~\cite{Kundu2013,Kundu2014}.

Some effects of sticky attractions have been characterized using simulations in Ref.~\cite{Longone2010}. 
On the one hand, critical temperatures $T_c^*$ (but no critical densities) of the gas--liquid transition have been estimated
from adsorption isotherms. $T_c^*$ was shown to increase from $\approx 0.7$ ($L=2$) to $\approx 1.3$ ($L=10$).
On the other hand, the demixing transition was investigated for $T^*=2...8$ (i.e., well above the
gas--liquid transition). As for purely hard--core rods, the demixing transition was shown to be second order and sets in for $L \ge 7$. 
The critical packing fractions
$\eta_c^{\rm demix}$ were determined using the Binder cumulant and finite--size scaling.
It was shown that $\eta_c^{\rm demix}(T^*)$  {\em increases} with decreasing $T^*$ for a fixed $L$, a somewhat surprising result, 
i.e. the isotropic phase becomes more stable with increasing attractions between the rods. 
Naively, one would expect the opposite, namely that the tendency to demix would 
increase with stronger attractions since the latter favor parallel alignment.    

\begin{figure}
 \begin{center}
 \includegraphics[scale=1]{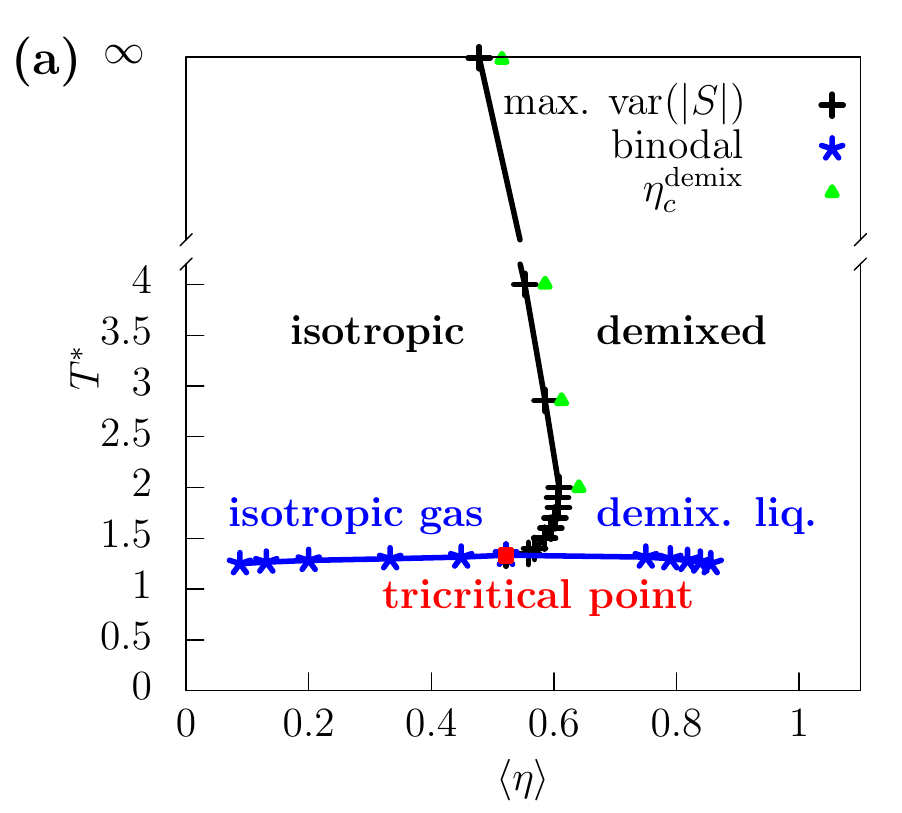}\hspace{0.4cm}
 \includegraphics[scale=0.75]{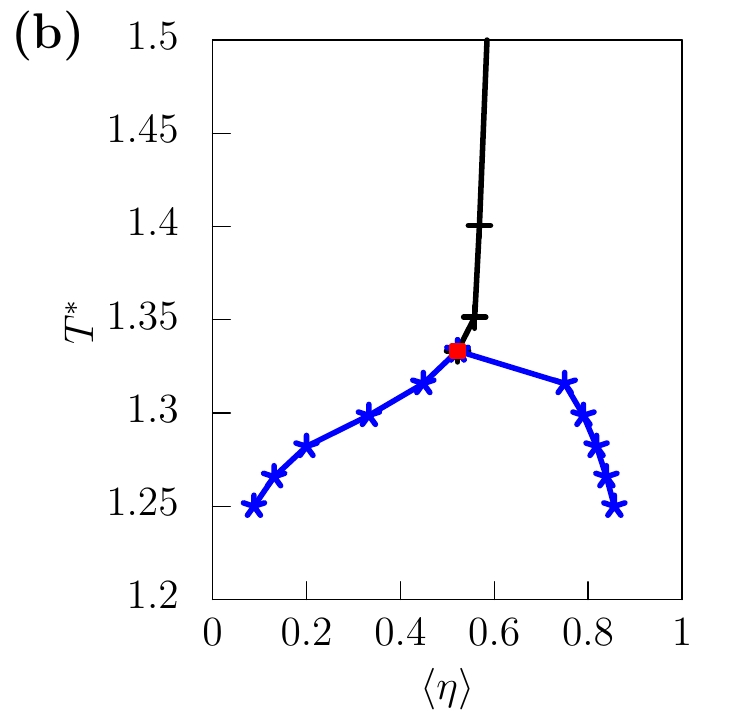}
  \end{center}
 \caption{Phase diagram for $L=10$ in 2D, in full in (a) and zoomed-in around the tricritical point in (b). Black symbols connected with lines show the demixing transition,
obtained from var($|S|$) and a system size $128^2$. 
 Green data points for the demixing transition are obtained by FSS in Ref.~\cite{Ramirez2008}. 
 The hard rod transition corresponds to $T^*=\infty$. Blue symbols connected with lines show the gas--liquid 
 binodal. 
  For the system size $128^2$, we estimate the location of the tricritical point at 
  $(T^*_c,\eta^{\text{demix}}_c) \approx (1.33, 0.52)$. }
 \label{fig:2DL10}
\end{figure}

These results leave open the question of the global phase diagram for $L \ge 7$. 
We have determined it for the exemplary
case of $L=10$. We locate the demixing packing fraction $\eta_c^{\rm demix}(T^*)$ at the maxima in var($|S|$), 
and the gas--liquid binodal points via successive umbrella sampling. The results are shown in Fig.~\ref{fig:2DL10}.

As can be seen in Fig.~\ref{fig:2DL10}, starting from $T^*= \infty$ and decreasing $T^*$ we reproduce the
increase in $\eta_c^{\rm demix}(T^*)$. There is a small shift towards lower packing fractions 
compared with the results of Ref.~\cite{Longone2010},
which is induced by the finite system size $M=128$ used here. Approaching $T_c^*$, the transition line $\eta_c^{\rm demix}(T^*)$
changes slope and terminates on the gas--liquid binodal. Thus the terminal point is most likely a tricritical
point. This entails (and is seen in our simulations) that below $T^*_c$ the coexisting liquid is in a demixed state.

Phase diagrams in 2D for sticky rods were investigated in Ref.~\cite{Mortazavifar2017} using density functional
theory (DFT) in the form of lattice fundamental measure theory (FMT). The FMT predicts the same topology of the phase diagram, i.e.
a line of second--order demixing transitions for decreasing temperatures, which end in a tricritical point.
However, there are serious quantitative discrepancies. The packing fractions $\eta_c^{\rm demix}(T^*)$ are much 
smaller in the FMT, and the tricritical point is at a much higher temperature. In simulations, the gas--liquid binodal
is very flat around $T^*_c$, similar to the behavior of simple liquids in 2D. The FMT renders a binodal that is considerably 
distorted toward higher temperatures. It also does not recover the unusual behavior of increasing $\eta_c^{\rm demix}(T^*)$ with 
decreasing $T^*$.

\begin{figure}
 \centerline{\includegraphics[width=5cm]{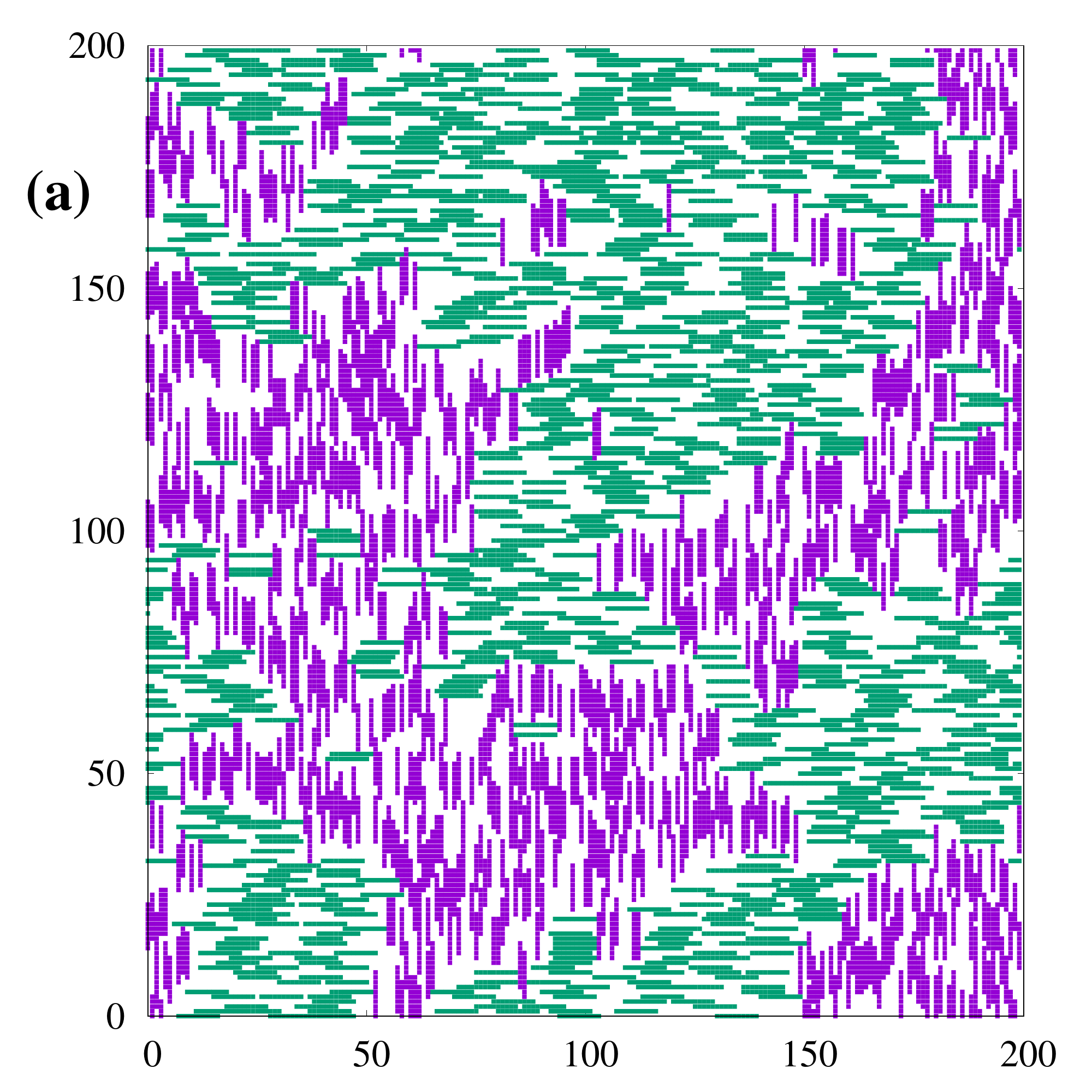} \includegraphics[width=5cm]{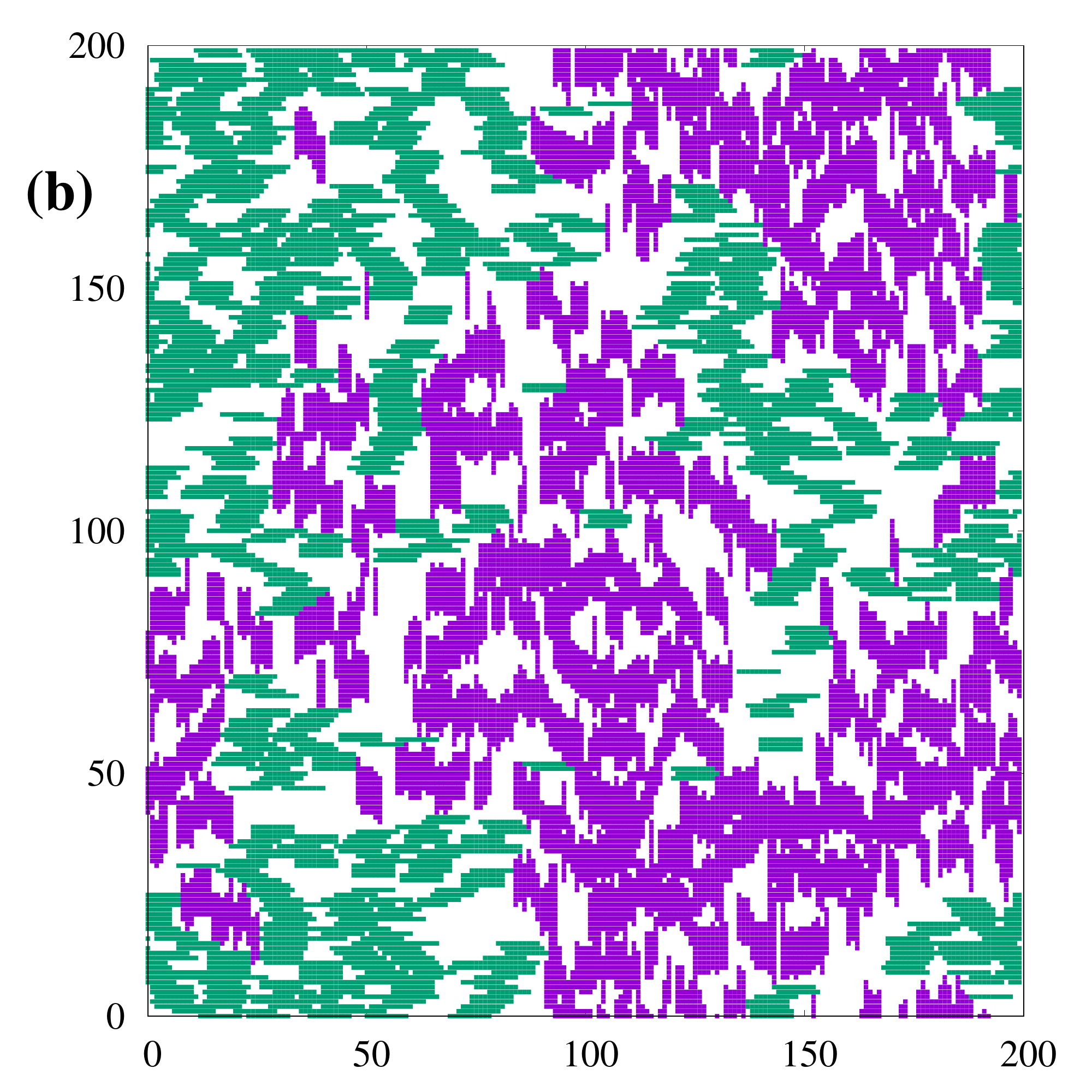}\includegraphics[width=5cm]{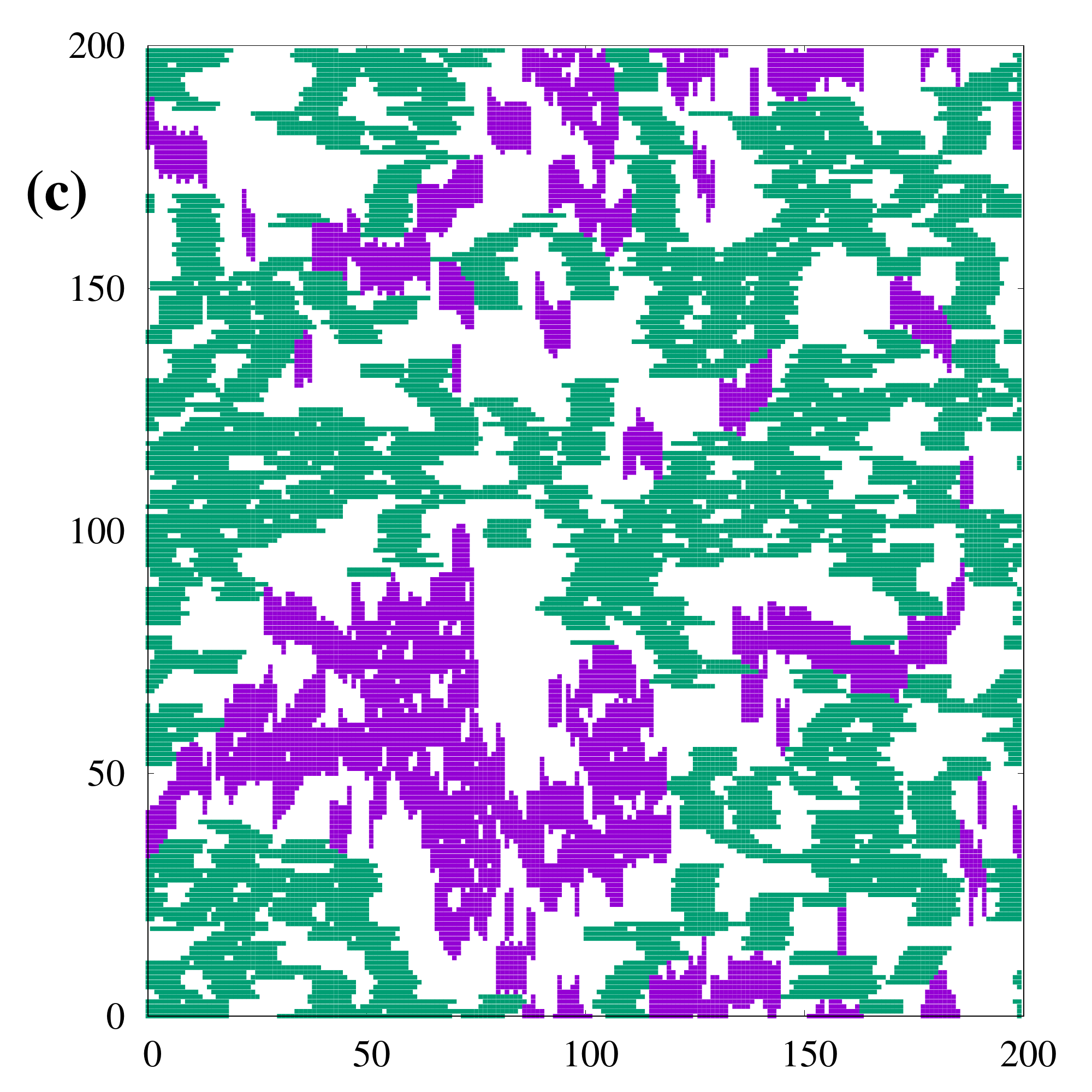}}
 \centerline{\includegraphics[width=5cm]{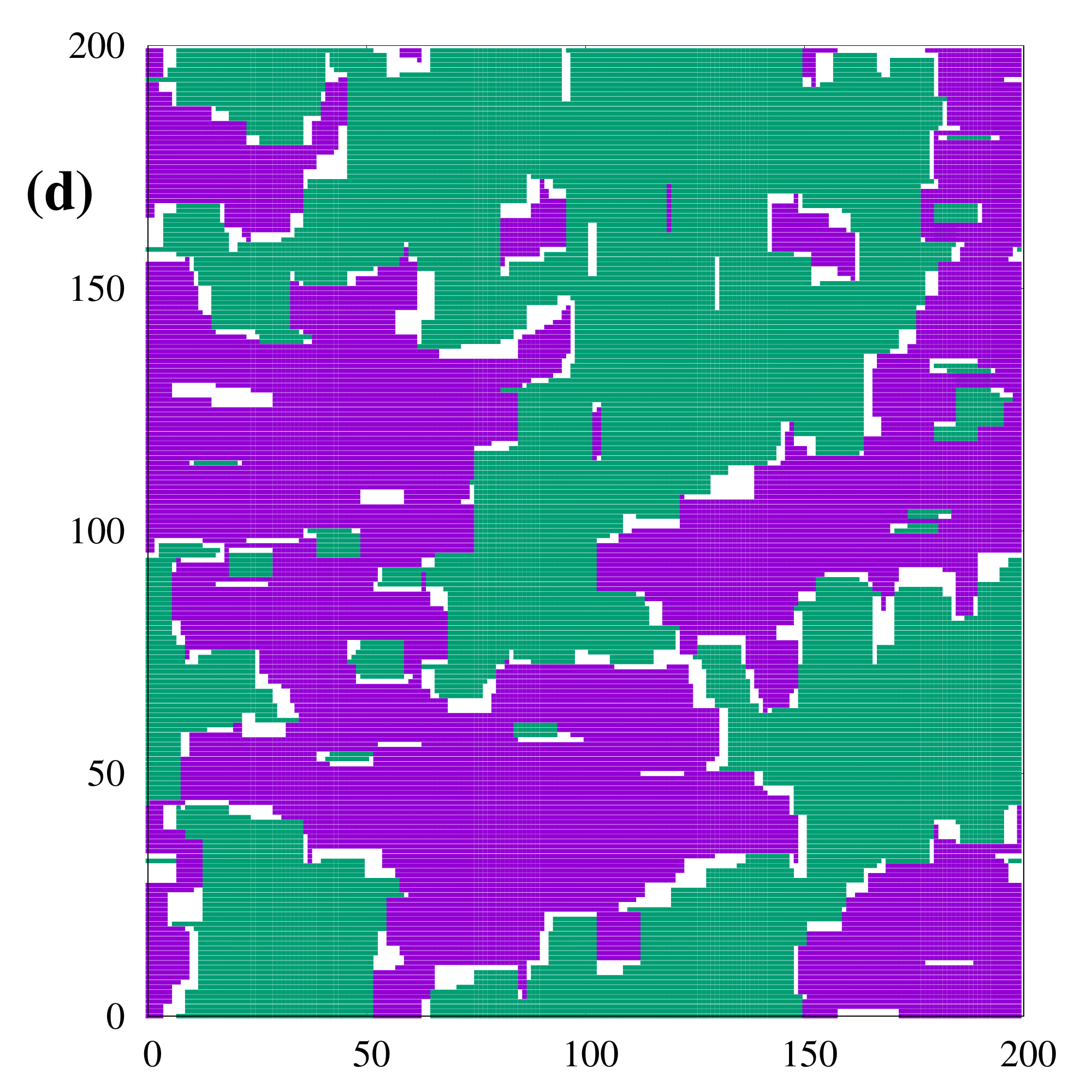} \includegraphics[width=5cm]{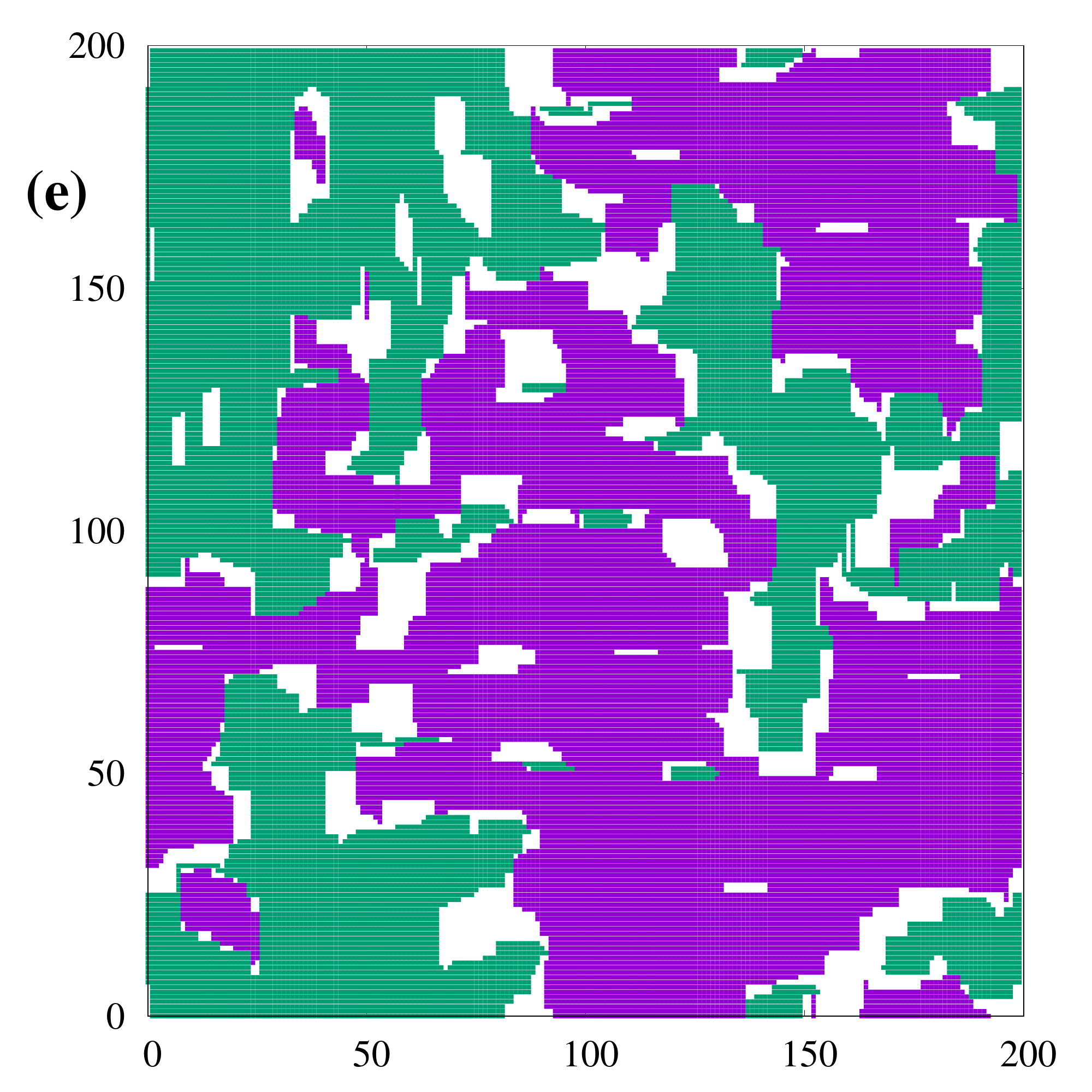}\includegraphics[width=5cm]{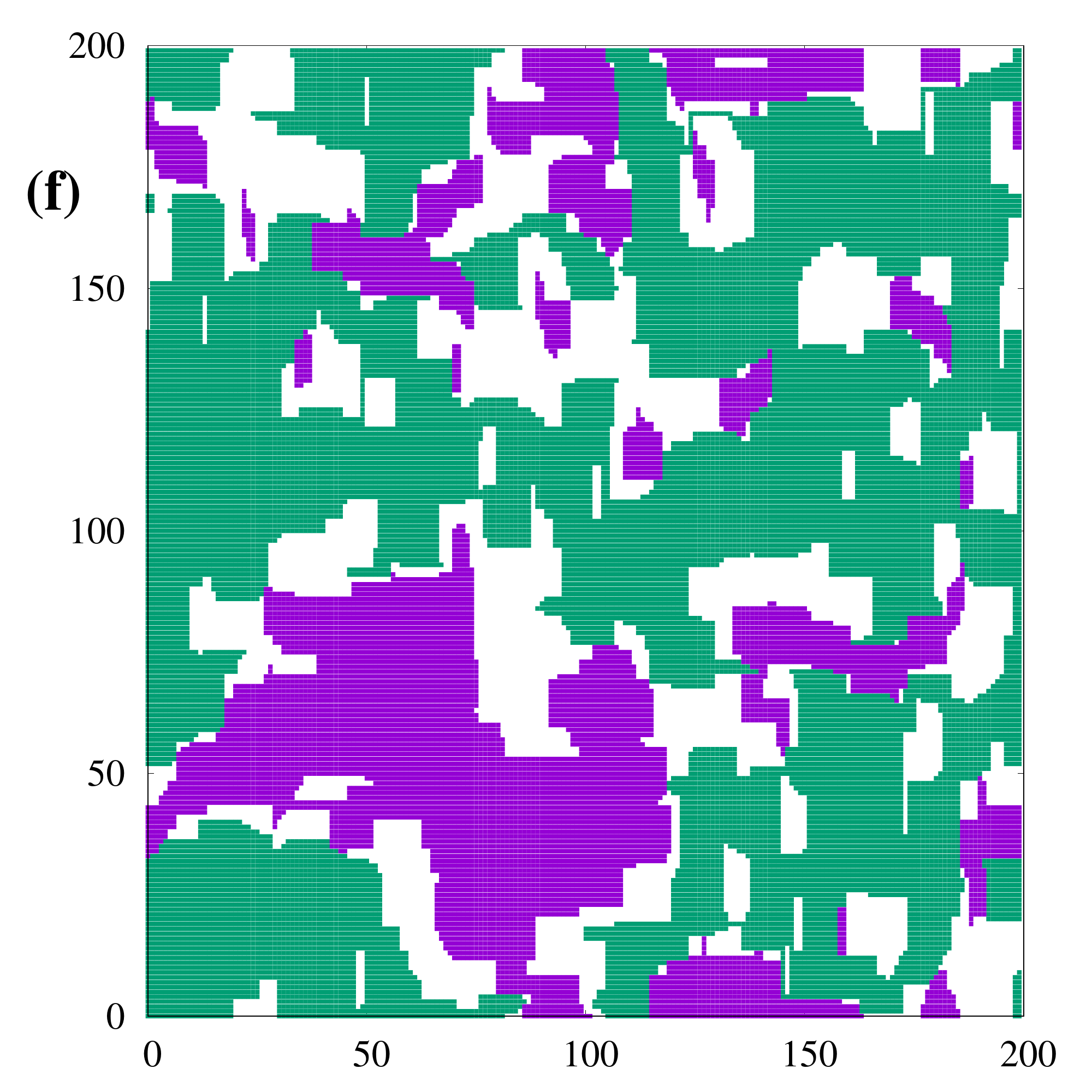} }
 \caption{Snapshots of rod occupation fields at three statepoints in the vicinity of
the demixing line.
Rod occupation fields $O_x$ and $O_y$ for (a) $T^*=\infty$, $z=0.24$ ($\langle \eta\rangle =0.48$),
(b)  $T^*=2$, $z=0.0022$ ($\langle \eta\rangle =0.59$) and
(c)  $T^*=1.5$, $z=0.0004$ ($\langle \eta\rangle =0.58$).
Corresponding snapshots for $O_X$ and $O_Y$ in (d), (e) and (f). 
The system size is $M^2 = 200^2$.}
 \label{fig:2Ddomains}
\end{figure}

All these discrepancies point to an important role played by larger--scale fluctuations. Simulation snapshots near
$\eta_c^{\rm demix}(T^*)$ show that the system splits into larger domains where the rods are oriented in either one or the other directions,
yet the order parameter $|S|$ remains small (see Fig.~\ref{fig:2Ddomains}(a)--(c)). 
At high temperatures,
these domains are loosely packed, but become increasingly dense for lower $T^*$ where attractions become more important.
To quantify these domains, we consider the orientation--specific binary occupancy fields $O_x(\vect s)$
and $O_y(\vect s)$ (which are 1 for lattice points covered by $x$--oriented resp. $y$--oriented  rods, 0 otherwise).
We define new binary occupancy fields $O_X(\vect s)$ and $O_Y(\vect s)$ by the following rule:
\bea
  O_X(\vect s) &=& \left\{ \begin{matrix} 1 && 
          \text{($\vect s$ cannot be covered by an $y$--oriented rod, given occupancy $O_x$)} \\
                   0 & & \text{(otherwise)} \end{matrix}   \right. \\
  O_Y(\vect s) &=& \left\{ \begin{matrix} 1 && 
          \text{($\vect s$ cannot be covered by an $x$--oriented rod, given occupancy $O_y$)} \\
                   0 & & \text{(otherwise)} \end{matrix}   \right. \nonumber
\eea  
In essence, $O_X$ defines more-or-less compact $X$--domains (belonging to $x$--oriented rods) whose points cannot be covered by a
 $y$--oriented rod. Likewise $O_Y$ defines $Y$--domains, see also Fig.~\ref{fig:2Ddomains}(d)--(f). One recognizes that the system is 
covered to a large extent by $X$-- and $Y$--domains with a few voids. One can define an order parameter
$S'$ by
\bea
  S' = \frac{ \eta_X - \eta_Y}{\eta_X + \eta_Y} \;,
\eea 
where $\eta_{X[Y]} = area(X[Y])/M^2$ is the packing fraction of the $X[Y]$ domain, with $M^2$ being the total area of the system. Interesting is that 
$|S'| \approx |S|$, i.e., $|S'|$ is an equivalent order parameter. A total
packing fraction $\eta_{XY}=\eta_X+\eta_Y$ of $X$-- and $Y$--domains is furthermore useful to define, as well as packing fractions of rods inside
these domains, i.e.  $\eta_{x,X}=LN_x / \text{area}(X)$ for $x$--oriented rods inside $X$--domains and $\eta_{y,Y}$ for $y$--oriented rods inside $Y$--domains, likewise. The total packing fraction is
$\eta = \eta_{x,X} \eta_X + \eta_{y,Y} \eta_Y$.

\begin{figure}
 \centerline{\includegraphics[width=7cm]{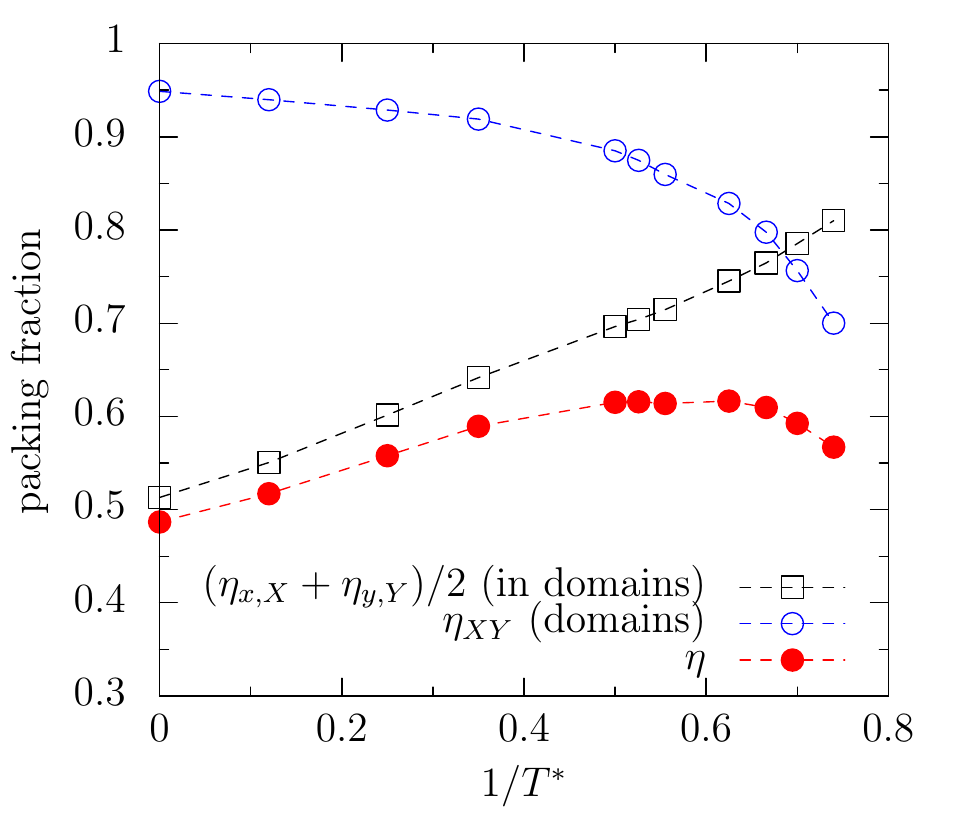} }
 \caption{ The behavior of the average packing fraction in the $X$[$Y$] domains, the packing fraction of the 
domains themselves and the rod packing fraction vs. attraction strength $1/T^*$ (for constant order parameter 
$|S|=0.5$, close to the demixing transition). Data are for a system size $M^2 = 200^2$. }
 \label{fig:eta_domains}
\end{figure}

The behavior of $\eta_c^{\rm demix}$ upon variation of temperature can now be rationalized. 
For zero attractions, $\eta_{XY}$ is rather high but the packing fractions inside the domains are moderate.
Furthermore $\eta_{x,X} \approx \eta_{y,Y}$.
With increasing attractions, $\eta_{XY}$ is only slowly
decreasing but $\eta_{x,X}$, $\eta_{y,Y}$ increase substantially.
This leads to the shift of the demixing transition to higher densities. At around $T^* \lesssim 2$,
$\eta_{XY}$ begins to decrease substantially, whence more voids appear in the system (see Figs.~\ref{fig:eta_domains} and \ref{fig:2Ddomains}(f)). This `signals' the approaching gas--liquid transition, and now $\eta_c^{\rm demix}$
decreases again.

\section{Three Dimensions}
\label{sec:3D}

Previous work has considered purely hard--core  rods ($T^*= \infty$) \cite{Gschwind2017,Rajesh2017}. For such hard rods with lengths $L \le 4$, 
the isotropic phase is stable up to packing fractions close to 1. For systems with rod--lengths $L=5$ and 6, a nematic transition
to a nematic$^-$ phase occurs at packing fractions $\eta^{\rm nem}\approx 0.874$ 
$(L=5)$ and $\eta^{\rm nem}\approx 0.68$ 
$(L=6)$.
In the lattice system, the nematic$^-$ phase is equivalent to layers filled with particles of the two majority species in a disordered fashion.
Particles of the third, minority species (oriented perpendicularly to these layers) occasionally pierce these layers. 
Systems of rods with lengths $L \ge 7$ entail a transition  between the isotropic and a nematic$^+$ phase. The transition packing fractions
are approximately $\propto 1/L$.   

The isotropic--nematic transition is presumably of very weak first order \cite{Gschwind2017}. A finite coexistence gap
(difference between packing fractions of the coexisting isotropic and nematic state) was not resolvable.
The transition is conspicuous as peaks in var($m$). In contrast, var($\eta$) shows only
a slight peak at the transition, embodying weak particle--number fluctuations at the transition.  

\subsection{$L=4$: gas--liquid transition only}

\begin{figure}
 \begin{center}
   \includegraphics[scale=0.75]{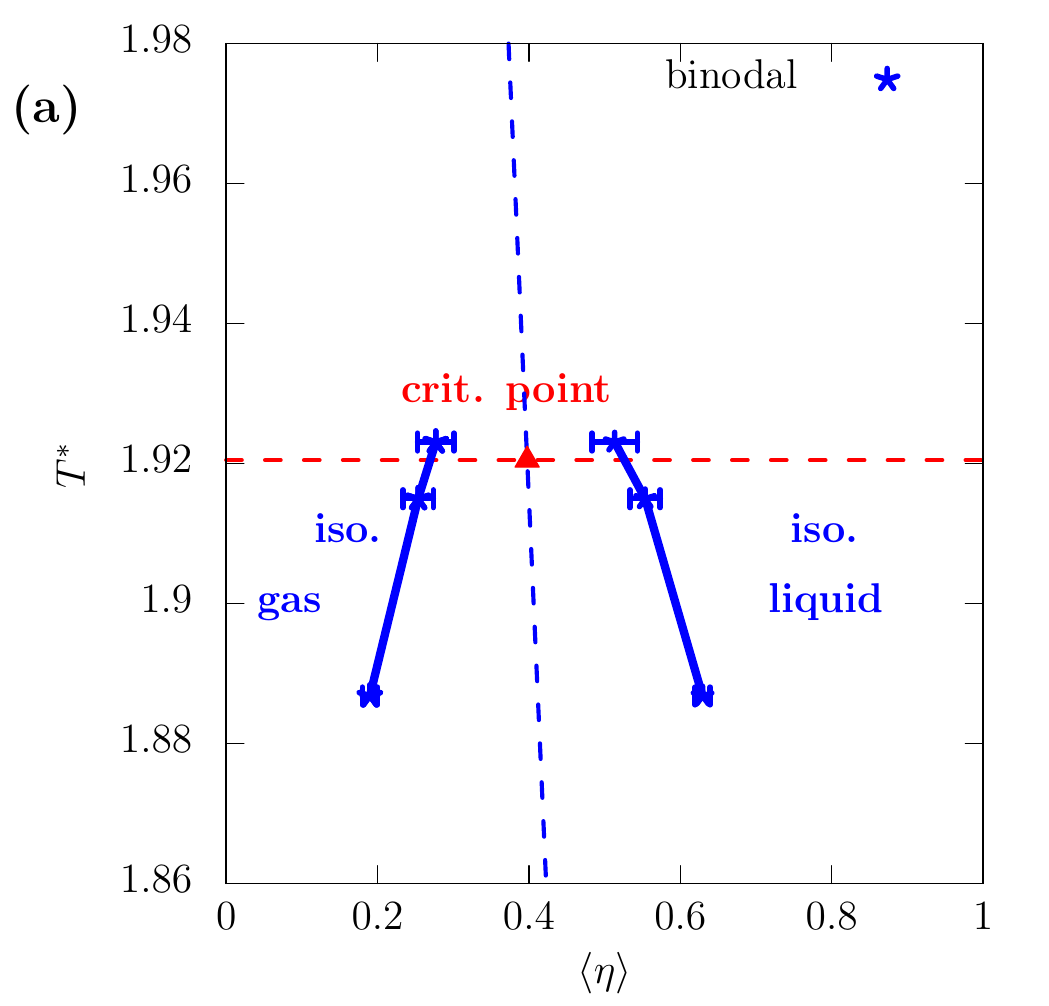} \hspace{1cm}
   \includegraphics[scale=0.75]{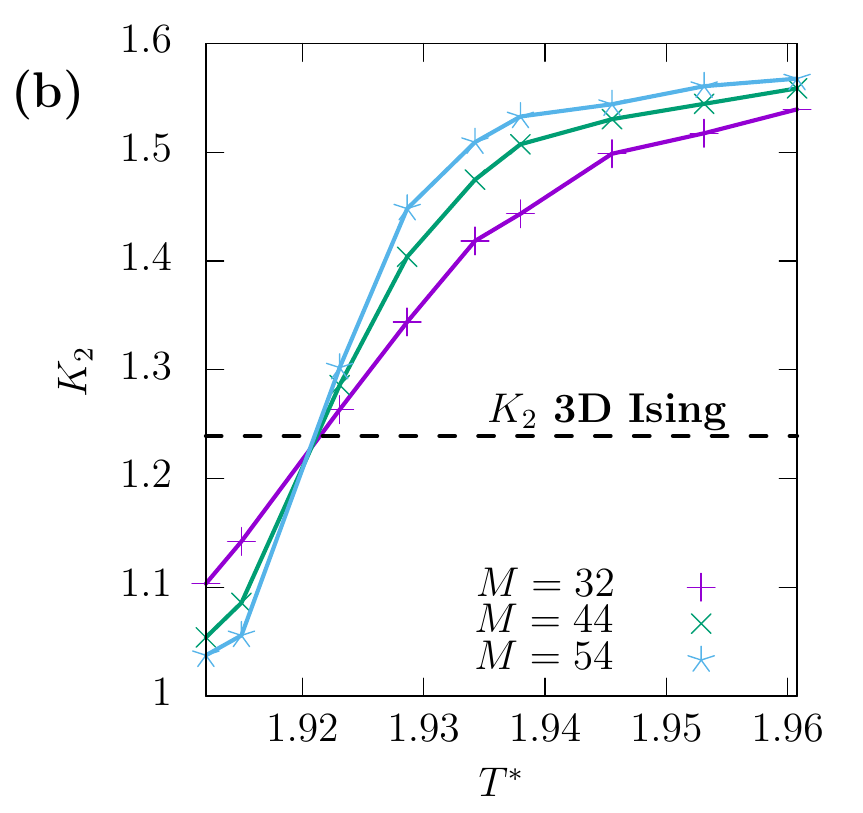} 
 \end{center}
 \caption{(a) Phase diagram for $L=4$ in 3D. Binodal points (blue symbols with error bars) are from 
the location of equal--area peaks in $P(\eta)$ and a system size
$M^3=32^3$. The error was estimated from the peak--fitting procedure.
The critical temperature $T_c^*$ was obtained by FSS (see (b)) and the critical packing fraction is at the
intersection of the line connecting the mean packing fraction of isothermal binodal points (blue dashed line) 
with horizontal line at $T_c^*$ (red dashed line). 
(b) Binder cumulants $K_2$ for three different system sizes for temperatures near $T_c^*$. The critical
point value of $K_2$ for the 3D Ising model is shown as a horizontal line.
}
 \label{fig:L4_M32_TVsEta}
\end{figure}

Similarly as in purely--hard case, 
no nematic transition could be detected in the range $T^*=1.85...\infty$ and $\eta=0.00...0.90$ for systems with rod-length $L=4$. Yet, these systems show a gas--liquid transition. For a system size $M^3=32^3$, we have determined binodal points through double peaks in $P(\eta)$ 
(see Fig.~\ref{fig:L4_M32_TVsEta}(a)).
The critical point was determined using FSS with the cumulant $K_2(\eta(z) - \langle \eta(z) \rangle)$,
where $z$ is the activity at phase coexistence
(see Fig.~\ref{fig:L4_M32_TVsEta}(b)). Here, the critical temperature $T_c^* \approx 1.92$ is 
the $T^*$--coordinate of the approximate crossing point of the $K_2$ curves. The corresponding critical packing
fraction $\eta_c \approx 0.40$ was obtained in Fig.~\ref{fig:L4_M32_TVsEta}(a) as the $\eta$--coordinate of the 
intersection of the horizontal line 
$T^*=T^*_c$ and the line connecting the mean of the coexisting liquid and gas packing fractions on the binodal, 
$\frac{1}{2} (\eta_{\text{l}}-\eta_{\text{g}})$. Owing to the small system size, the histogram $P(\eta)$ shows two
distinct peaks for $T^*>T^*_c$, as well.

\subsection{$L=5$ and $6$: gas--liquid and nematic$^-$ transition} 

\begin{figure}
 \centerline{\includegraphics[scale=1]{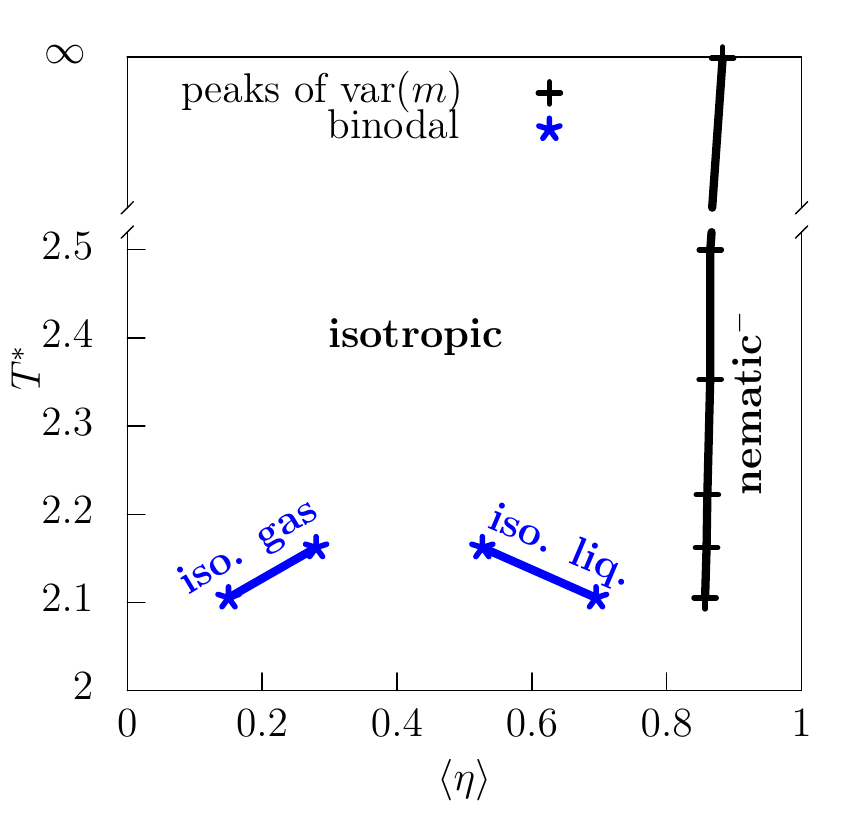}} 
 \caption{Phase diagram for $L=5$ in 3D for a system size $M^3=64^3$. The hard rod transition corresponds to 
$T^*=\infty$ and is located at a  packing fraction $\eta_c^\text{nem} \approx 0.88$
(in agreement with Refs.~\cite{Gschwind2017} and \cite{Rajesh2017}).
}
 \label{fig:L5_M64_TVsEta}
\end{figure}

Systems with rod-length $L=5$ show
an isotropic--nematic$^-$ transition that persists for all investigated temperatures (down to $T^* = 2.1$). The corresponding
transition packing fractions $\eta^{\rm nem}(T^*)$ become smaller with decreasing temperature, but,  depend on  $T^*$ 
only weakly, overall. 
The weak--first--order character changes very little with decreasing temperatures, as can be seen in the value of
var($\eta$) at the transition. The latter increases from $1\cdot 10^{-6}$ $(T^* = \infty)$ to
$8 \cdot 10^{-6}$ for $T^* = 2.1$, which points to a very small coexistence gap (see also below for estimates
of the coexistence gap from var($\eta$)).
Additionally, there is a gas--liquid transition with a critical temperature above $T^*= 2.1$. We have determined 
coexisting states (isotropic gas and isotropic liquid) using SUS down to $T^* = 2.1$. Unfortunately, for temperatures
further below we encountered
equilibration problems. 
Therefore, we can only speculate that the line of isotropic--nematic transitions ends on the binodal,
far ``right'' on the liquid side. Since the isotropic--nematic transition stays very weakly first order, such an end point
would amount to being a \emph{pseudocritical} end-point, which is presumably hard to resolve in simulations.     

\begin{figure}
\centering
  \includegraphics[scale=0.8]{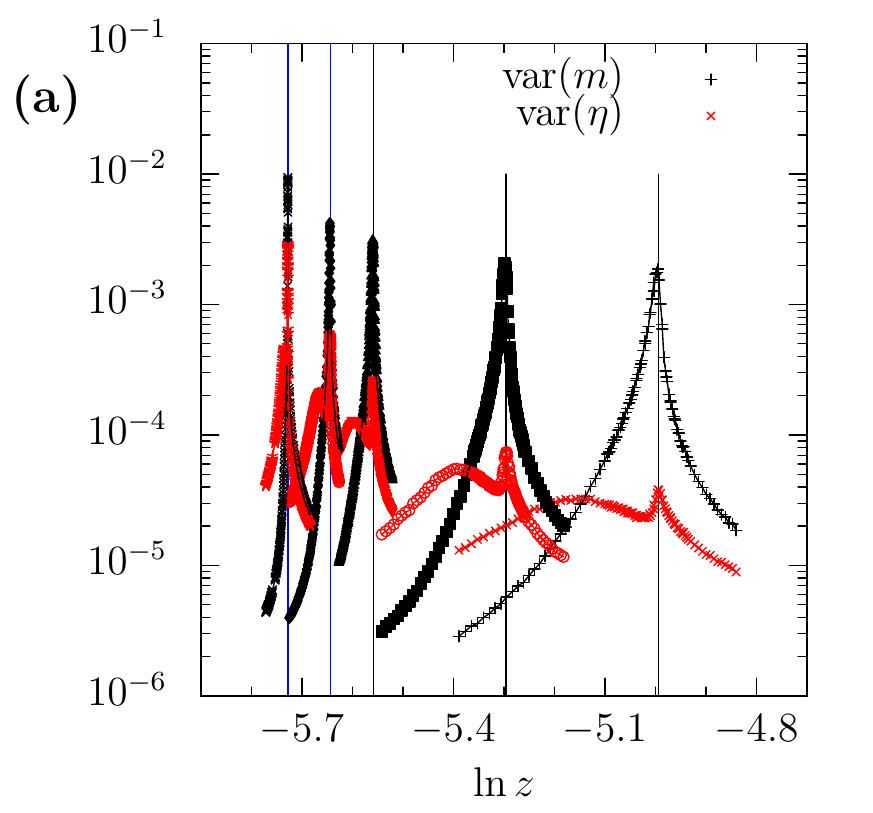}\hspace{1cm}
  \includegraphics[scale=0.8]{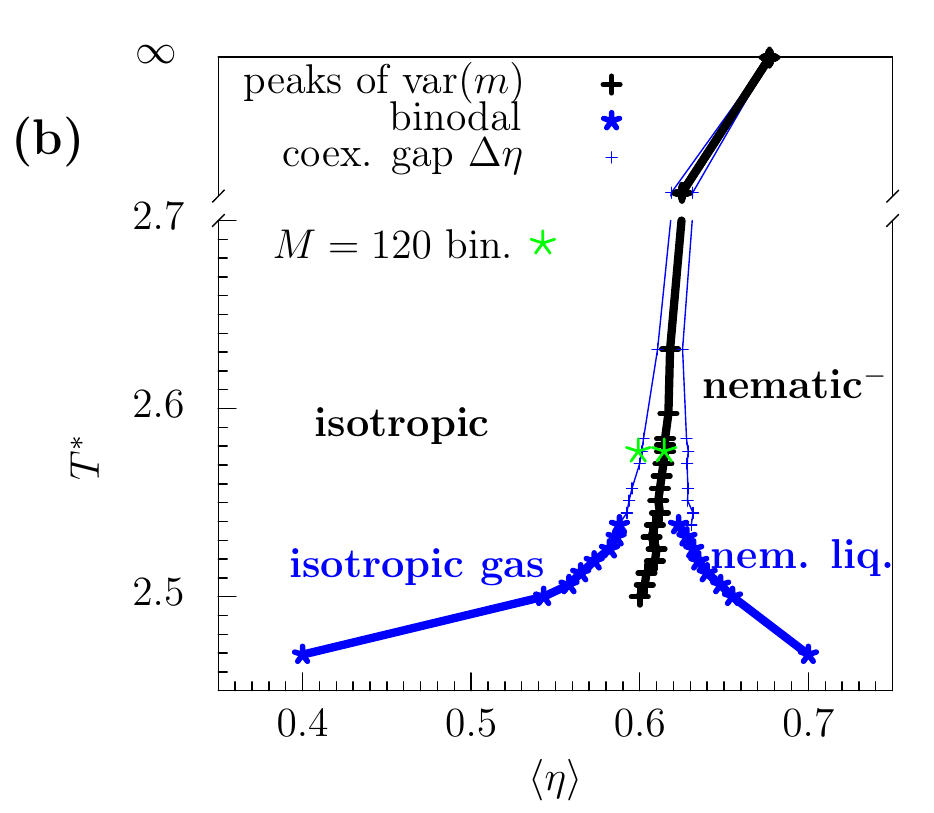}
 \caption{ 
(a) var($m$) and var($\eta$) isotherms as a function of $\ln z$ for a system size $M^3=64^3$. 
The temperatures (for pairs of curves from left to right) are $T^*=2.500$, $2.538$, $2.577$, $2.703$ and $2.857$. 
Vertical lines indicate $\ln z$ where var($m$) and var($\eta$) are maximal.
Blue vertical lines additionally indicate  that at this coexistence activity two peaks in $P(\eta)$ were distinguishable.
(b) phase diagram for $L=6$ in 3D for a system size $M^3=64^3$. 
Binodal points connected by thick blue lines were determined using histograms $P(\eta)$ that showed two
clearly--resolvable peaks. SUS was used at the lowest temperature. An upper limit for the coexistence gap
(binodal points connected by thin blue lines) 
were estimated from var($\eta$), see text. Green stars are binodal points from two peaks in $P(\eta)$
at $T^*=2.577$ for a large system size ($M^3=120^3$). 
}
  \label{fig:L6_M64}
\end{figure}

Systems with rod-length $L=6$ show an isotropic--nematic$^-$ transition which shifts toward lower packing fractions
as $T^*$ decreases (similarly to $L=5$). In contrast to the case of $L=5$,
we observe particle--number fluctuations increasing substantially with decreasing $T^*$. 
In Fig.~\ref{fig:L6_M64}(a) we show five isotherms for var($m(z)$) and var($\eta(z)$) in the temperature interval
$T^*=2.9...2.5$. Peak locations of var($m$) and var($\eta$) agree and signify the nematic transition in harmony.    
Curves of var($\eta(z)$) show a broad background signal with a sharp peak whose maximum increases from 
$\approx 3 \cdot 10^{-5}$ at $T^*=2.857$ to $\approx 2 \cdot 10^{-3}$ at $T^*=2.5$. Away from the sharp peak
(in the background),
the corresponding histogram $P(\eta)$ is described by a single Gaussian peak with variance $\sigma$.
The grand--canonical fluctuation relation for particle numbers in a finite system with volume $V$ expresses that
\bea
 \text{var}(\eta) = \frac{L}{V\beta} \langle \eta \rangle \chi = \sigma^2\;,
\eea 
where $\chi=[\eta_i(\partial p/\partial \eta_i)]^{-1}$ is the isothermal compressibility.
The background noise displays a broad, secondary peak, which corresponds to a maximum in the
compressibility $\chi$, and can be viewed as the supercritical `precursor' of the gas--liquid transition.
The nematic transition contributes to var($\eta$) at the position of the sharp peak. 
At the two lowest temperatures in Fig.~\ref{fig:L6_M64}(a) ($T^*=2.500$ and 2.538) the
corresponding histogram $P(\eta)$ displays clear double--peaks that are characteristic of the first--order nature of the transition.
The peaks are approximately  
described by Gaussians $\exp[ -(\eta-\eta_i)^2/(2\sigma_i^2)]$ where
$\eta_i$ is the peak position (coexistence packing fraction) and $\sigma_i$ a peak width.
At coexistence, contributions to var($\eta$) arise from the compressibilities $\chi_i$ of the bulk states, 
from switches between the coexisting bulk states and lastly from additional fluctuations at the interface between these.
At low temperatures, these latter fluctuations are small and var($\eta$) becomes:
\bea
  \label{eq:vareta_coex}
  \text{var}(\eta)& =& \frac{1}{4}(\eta_1 - \eta_2)^2 + \frac{L}{2V\beta}( \langle \eta_1 \rangle \chi_1  + 
    \langle \eta_2 \rangle \chi_2 ) \\ 
                  & \approx & \frac{1}{4}(\eta_1 - \eta_2)^2 + \frac{1}{2}( \sigma_1^2 + \sigma_2^2)\;. \nonumber
\eea   
The extracted coexistence gap $|\eta_1 - \eta_2|$ from Eq.~(\ref{eq:vareta_coex}) agrees well with the
coexistence gap determined from the peak locations in $P(\eta)$ for low temperatures.
We could not identify two well--separated peaks in $P(\eta)$ anymore for $T^* \gtrsim 2.54$ .
Thus, there must also be non--negligible interface fluctuations in play  contributing to var($\eta$) .
Nevertheless, we can estimate an upper limit from Eq.~(\ref{eq:vareta_coex}) for the coexistence
gap $|\eta_1-\eta_2|$  from the excess (over--background) height of the peak in var($\eta(z)$) at $z=z_\text{coex}$, as
the background is described by the second term on the right-hand-side of Eq.~(\ref{eq:vareta_coex}). 
We assumed that the binodal with this coexistence gap is situated
symmetrically around $\langle \eta \rangle(z_\text{coex})$.

The resulting phase diagram is shown in Fig.~\ref{fig:L6_M64}(b) (with binodal points both from double--peaks in $P(\eta)$ and from the
construction using var($\eta$)).
Upon decreasing the temperature,
we observe a gradual opening of the coexistence gap. In a rather small temperature window ($T^*=2.6...2.5$)
the coexistence gap widens rapidly. The binodal describes coexistence of a low--density isotropic
gas and a high--density nematic$^-$ liquid. 
As described, resolving the coexistence gap widening is not easy owing to the strong fluctuations in the system.

\begin{figure}
\centering
  \includegraphics[scale=0.8]{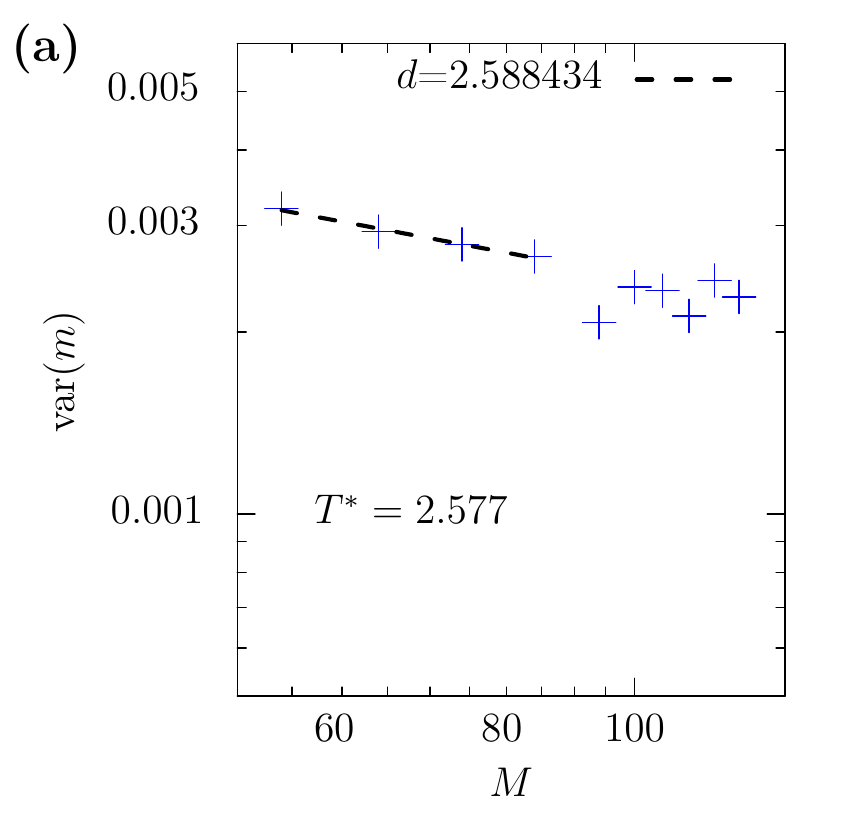}\hspace{1cm} 
  \includegraphics[scale=0.8]{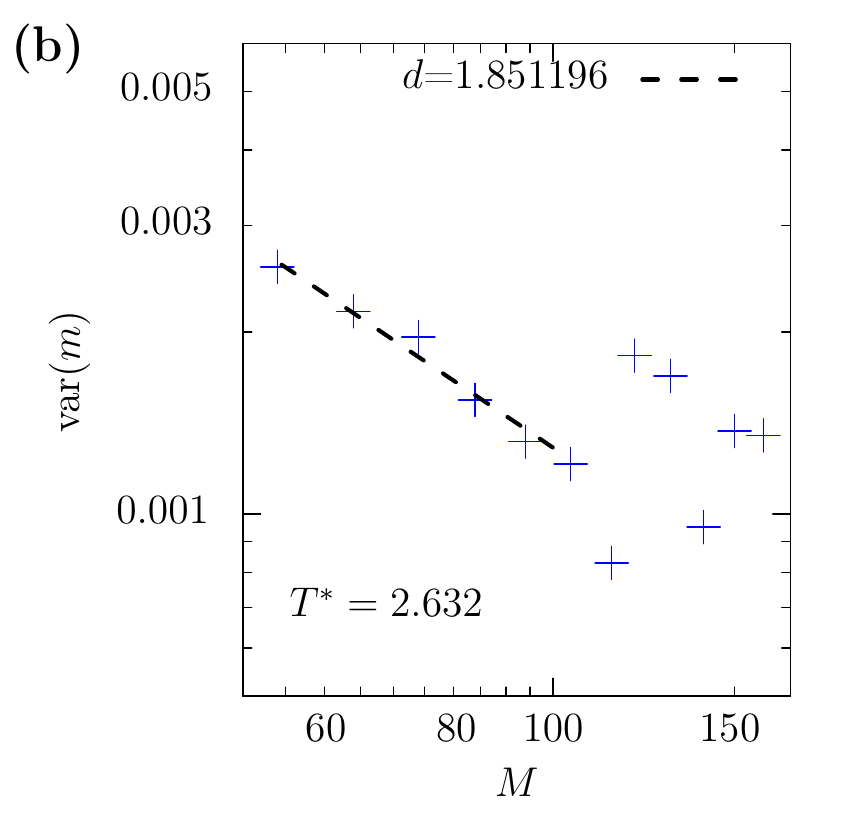} 
  \caption{ 
var($m$) vs. linear system size $M$ for (a) $T^*=2.577$, $z_{\text{coex}} = 3.850\cdot 10^{-3}$ and
(b) $T^*=2.632$, $\ln z_{\text{coex}} = 4.315\cdot 10^{-3}$. For smaller $M$, data can be nicely fitted by
$\ln \text{var}(m) =(d-3) \ln M + \text{const.}$ (dashed lines).
}
  \label{fig:fss}
\end{figure}

The assumption of a first--order character of the nematic transition for all temperatures is reasonable
given (i) the symmetry of the model, (ii) the established weak first--order character for hard rods ($T^* \to \infty$)
\cite{Gschwind2017} and (iii) corresponding results for continuum models. Nevertheless, we investigated 
the finite--size behavior of the system more closely at two temperatures around $T^* = 2.6$.
In Fig.~\ref{fig:fss}, the behavior of var($m$) as a function of linear system size $M$ is shown at the
two temperatures $T^*=2.577$ and $T^*=2.632$, at each of their coexistence activities $z_\text{coex}$. 
One may attempt to fit the data with a power law 
$\text{var}(m) \propto M^{d-3}$. For a first order transition, $d=3$ in three dimensions.
At the lower temperature $T^*=2.577$, we recognized two peaks in the histogram 
$P(\eta)$ for a system size $M=120$ (green stars in Fig.~\ref{fig:L6_M64}(b)). This is consistent with
the behavior of var($m$) in the double--logarithmic plot in  Fig.~\ref{fig:fss}(a), which indicates a
change of slope at $M \lesssim 100$. A double peak in $P(\eta)$ could not 
be resolved at the higher temperature $T^*=2.632$. Consequently, the slope $d \neq 3 $ up until $M \gtrsim 100$ in Fig.~\ref{fig:fss}(b), indicating
critical behavior, which, however, changes for larger $M$. Unfortunately, a fit would not be reliable for larger $M$ owing to the small number of fully de--correlated Monte-Carlo samples.   
 
\begin{figure}
 \centerline{\includegraphics[width=7cm]{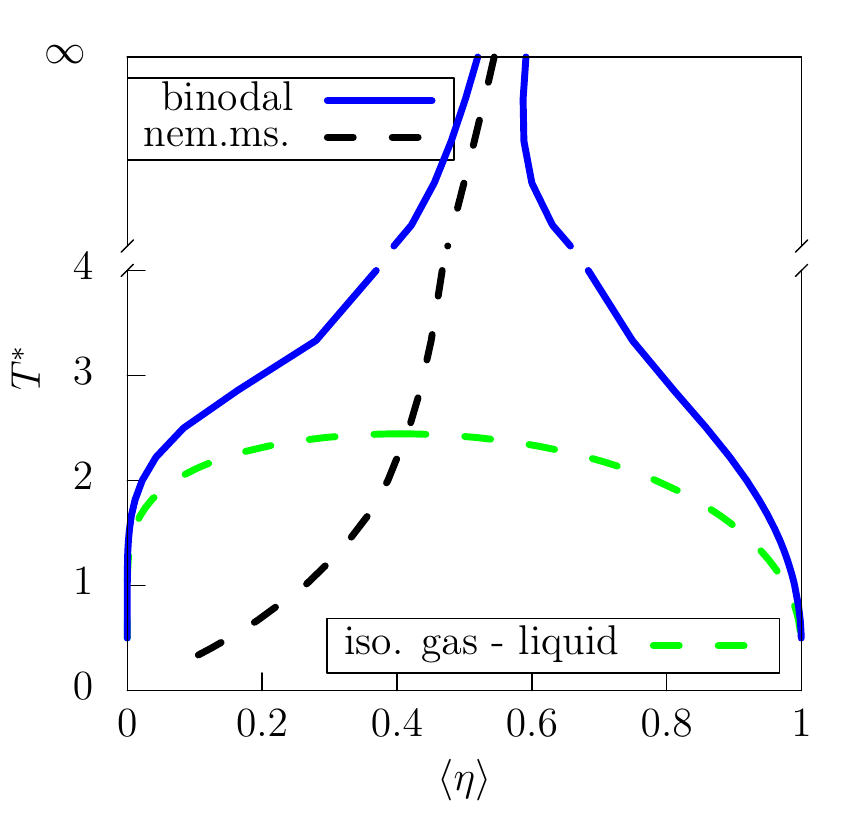}} 
 \caption{
Phase diagram for $L=6$ in 3D from FMT (data taken from Ref.~\cite{Mortazavifar2017}). Solid lines are
binodals for the transition between isotropic and nematic$^+$ states. The green dashed line is a binodal between
isotropic gas and isotropic liquid. The black dashed lines marks the onset of metastability for the nematic phase.  }
 \label{fig:L=6fmt}
\end{figure}

The phase behavior is in  contrast, quantitatively, to existing theoretical treatments of the isotropic--nematic transition
both in lattice and continuum systems. For the lattice system, the 
phase diagram from the FMT functional derived in Ref.~\cite{Mortazavifar2017} is reproduced in Fig.~\ref{fig:L=6fmt}.
For hard rods, the FMT predicts a strong first order transition from an isotropic to a nematic$^+$ state with a coexistence
gaps $\Delta\eta \approx 0.07$. With increasing attractions (lower $T^*$), the coexistence gap significantly widens
already at high temperatures
and  crosses smoothly over to a transition between a thin isotropic gas and a dense nematic$^+$ liquid
(thick black line in Fig.~\ref{fig:L=6fmt}). Additionally, we show the isotropic gas--liquid binodal (green dashed line in 
Fig.~\ref{fig:L=6fmt}, obtained by a restricted minimization of the FMT functional with $m=0$) and the onset of metastability
of the nematic phase (\tr{black} dot--dashed line in Fig.~\ref{fig:L=6fmt}). The qualitative behavior of these lines
is strikingly similar to the phase diagram obtained by simulations, the gas--liquid critical temperatures $T_c^*$
are close, but the critical packing fraction $\eta_c$ differs.  

\subsection{$L=8$: gas--liquid and nematic$^+$ transition}

\begin{figure}
\centering
  \includegraphics[scale=1]{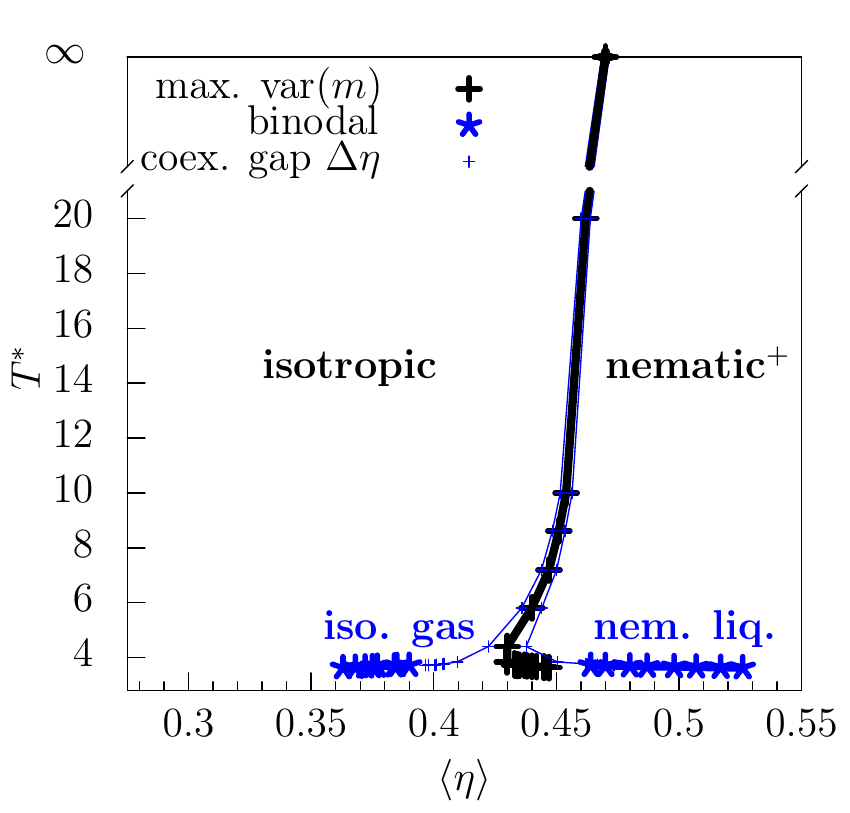}
\caption{Phase diagram for $L=8$ in 3D for a system size $64^3$. 
Binodal points connected by
thick blue lines were determined using histograms $P(\eta)$ which showed two clearly
resolvable peaks. Binodal points connected by
thin blue lines were estimated from var($\eta$) as for $L=6$.
}
 \label{fig:L8_M64}
\end{figure}

The hard--rod system with rod-length $L=8$ shows an isotropic-nematic$^+$ transition which also persists for lower temperatures.
As for $L=6$,  the transition widens and becomes
 one between an isotropic gas and a nematic liquid within a rather short temperature interval (here at around $T^*=3.721$). As before, we pinpoint binodal points for 
$T^* \lesssim 3.721$ through the locations of the two peaks in $P(\eta)$. For $T^* \gtrsim 3.721$ we obtain them from 
var($\eta$) as an upper limit by Eq.~(\ref{eq:vareta_coex}). The behavior of var($\eta(z)$) and var($m(z)$) is very similar to the case of $L=6$,
see Fig.~\ref{fig:L6_M64}. The resulting phase diagram is shown in Fig.~\ref{fig:L8_M64}.
 
\begin{figure}
\centering
   \includegraphics[width=7cm]{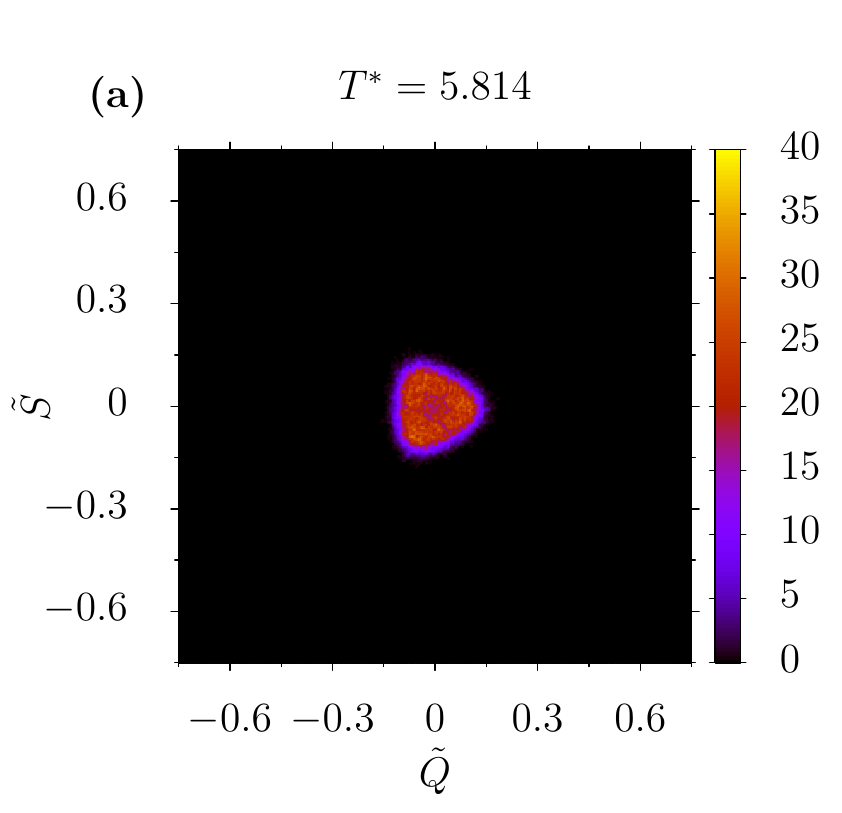}\hspace{1cm}
  \includegraphics[width=7cm]{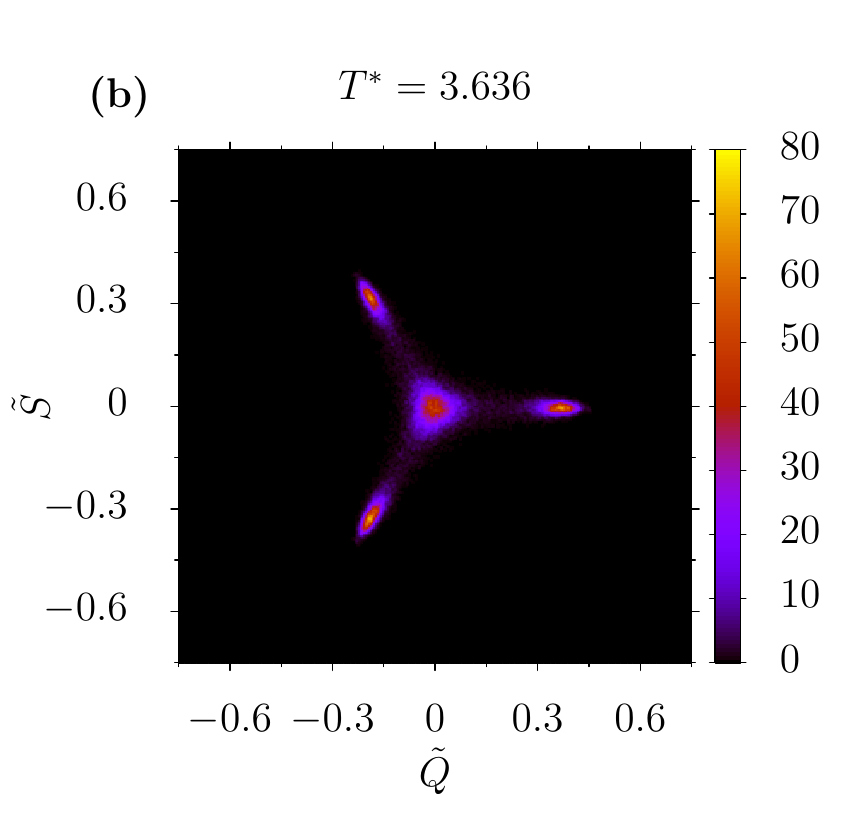}
\caption{$\tilde{Q}$-$\tilde{S}$ histograms taken at the isotropic-nematic$^+$ transition for the temperatures
(a) $T^{*}=5.814$, showing a \emph{weakly} first--order, and (b) $T^{*}=3.636$, a \emph{clearly} first--order transition for $L=8$ in 3D. 
}
 \label{fig:heatMapQSL8zcoex}
\end{figure}

We illustrate the behavior of the system at coexistence for one state above $T^*= 3.721$ 
(where we have a weak first--order transition) and one state below $T^*= 3.721$ (where we have 
a strong first--order transition). 
Histograms in the $\tilde Q$--$\tilde S$ order parameter plane are shown in Fig.~\ref{fig:heatMapQSL8zcoex}(a)
for $T^*=5.814$ and  Fig.~\ref{fig:heatMapQSL8zcoex}(b) for $T^*=3.636$. At the higher temperature, two peaks in
$P(\eta)$ were not discernible. This translates into a broad distribution in the $\tilde Q$--$\tilde S$ order--parameter plane. 
Unfortunately, this precludes an application of SUS methods on $\tilde Q$--$\tilde S$--histograms to extract
the coexisting packing fractions. In contrast, we see a clear superposition of isotropic and
nematic peaks at the lower temperature, i.e. a superposition of the first two histograms in the schematic Fig.~\ref{fig:scheme}.

\section{Summary and conclusion}
\label{sec:summary}

We have studied phase diagrams of a simple model for rods with short--ranged attractions on square and cubic
lattices using grand--canonical Monte Carlo simulations. This model may be viewed as the simplest lattice realization
of a model for anisotropic particles with competing gas--liquid and ordering transitions. 
The phase diagrams show qualitative and
quantitative differences to those of simple continuum models (e.g. spherocylinders with short--ranged attractions). 

In 3D and for weak attractions, isotropic--nematic transitions are of weakly first order with a small
coexistence gap (not resolvable in particle number histograms), owing to large fluctuations of nematic
subdomains. Their coming into being is favored by the discreteness (restriction) of the rod orientations on the lattice.
For rod--lengths $L=5$ and 6, the ordered phase is a nematic$^-$ phase with two majority species,
whereas for higher $L$, it is a nematic$^+$ phase with one majority species.
For $L=6$ and 8 and increasing attractions (decreasing temperature), the weak--first--order character persists down to
a length--specific transition temperature. There is a small window around this temperature where the coexistence gap quickly
widens, upon which the isotropic--nematic transition becomes strongly first order between an isotropic gas and a nematic liquid. 
If studied with small systems, the
phase diagrams appears to show a line of critical isotropic--nematic transitions hitting the gas--liquid binodal   
at a tricritical point. This is a bit surprising in view of results from Landau theory and simulation
for continuum models. Systems with $L=5$ entail a transition between an isotropic gas and an isotropic 
liquid. The isotropic--nematic transition occurs at much higher packing fractions and does not `disturb' the
isotropic gas--liquid transition near its critical point.

We have studied the 2D system in the exemplary case of $L=10$.
On the square lattice, a transition in the nematic order is realized by the two possible
rod orientations demixing.
Here, increasing attractions initially stabilize the
isotropic phase and shift the second--order demixing transition to higher packing fractions.
We rationalized this behavior via an analysis of the structure of the configurations, which shows domains  dominated effectively by only one species.
Increasing attractions have the effect of densifying these domains without changing the demixing
order parameter. However, with stronger attractions, the  
gas--liquid transition `nearby' seems to make the demixing transition shift to lower packing
fractions: the domains are compactified, but, shrink in size and thus make room for voids. 
The demixing line hits the gas--liquid binodal at  a presumably tricritical point.  

Phase diagrams of this system calculated from density functional theory (lattice fundamental measure theory) are
qualitatively correct but show significant quantitative differences. It appears that the fluctuations
in nematic domain size and their distribution are not captured correctly.  

\noindent
\textbf{Outlook:}\quad
We would like to extend the notion in our claim that lattice models of hard rods with sticky attractions are a worthwhile complement and counterpart to existing literature on continuum models. We suggest a few similar model systems: Hard rods with non-unit width \cite{Kundu2014} and rods on triangular lattice \cite{Kundu2013} have only been studied in the hard--core case.  Hard, polydisperse rods were studied in a random-sequential-adsorption-setting in Ref. \cite{Hart2016}, but it seems the equilibrium phases have not been determined even for the hard--core case. 
On another note, we see potential in using modern statistical analysis methods from information theory such as in  Ref. \cite{Ramirez-Pastor2017}, who use data compression. Machine learning can be applied, which is an approach we took and will demonstrate in an upcoming publication.

\acknowledgments
This work is supported in part by the DFG/FNR INTER project 
``Thin Film Growth" by the Deutsche Forschungsgemeinschaft (DFG), project OE 285/3-1.
The authors acknowledge support by the state of Baden-W\"urttemberg through bwHPC for computational resources.

\bibliographystyle{PRE}
\bibliography{2019_latticeRods}

\end{document}